\documentclass[onecolumn,english,smallextended]{svjour3}
\usepackage{graphicx}
\usepackage{textcomp}
\usepackage{amssymb}

\makeatletter

\smartqed  

\usepackage{amsfonts}

\makeatother

\usepackage{babel}

\begin{document}

\title{On the geometrization of matter by exotic smoothness}

\author{\author{Torsten Asselmeyer-Maluga \and Helge Ros{\'e}}
\institute{T. Asselmeyer-Maluga \at German Aerospace Center (DLR), Rutherfordstr. 2, 12489 Berlin, Germany \\ \email{torsten.asselmeyer-maluga@dlr.de} \and H. Ros{\'e} \at Fraunhofer FIRST, Kekul{\'e}str. 7, 12489 Berlin, Germany\\ \email{helge.rose@first.fraunhofer.de} }
\date{Received: date / Accepted: date}}
\maketitle
\begin{abstract}
In this paper we discuss the question how matter may emerge from space.
For that purpose we consider the smoothness structure of spacetime
as underlying structure for a geometrical model of matter. For a large
class of compact 4-manifolds, the elliptic surfaces, one is able to
apply the knot surgery of Fintushel and Stern to change the smoothness
structure. The influence of this surgery to the Einstein-Hilbert action
is discussed. Using the Weierstrass representation\emph{,} we are
able to show that the knotted torus used in knot surgery is represented
by a spinor fulfilling the Dirac equation and leading to a Dirac term
in the Einstein-Hilbert action. For sufficient complicated links and
knots, there are ''connecting tubes'' (graph manifolds, torus bundles)
which introduce an action term of a gauge field. Both terms are genuinely
geometrical and characterized by the mean curvature of the components.
We also discuss the gauge group of the theory to be $U(1)\times SU(2)\times SU(3)$.
\end{abstract}

\section{Introduction: A geometrical model of matter}

The proposal to derive matter from space was considered by Clifford
as well by Einstein, Eddington, Schr{\"o}dinger and Wheeler with
only partly success (see \cite{MiThWh:73}). In a recent overview,
Giulini \cite{Giulini09} discussed the status of geometrodynamics
in establishing particle properties like spin from spacetime by using
special solutions of general relativity. Similar ideas are discussed
in the model of Bilson-Thompson \cite{Bilson-Thompson2005}, in its
loop theoretic extension \cite{BTMarkSmolin2007} and in a model of
Finkelstein \cite{Finkelstein2009} using the representation of knots
by quantum groups. These approaches are using generalized geometric
structures, special solutions of general relativity or a larger class
of connections in some vector bundles. In this paper we will consider
the smoothness structure of a 4-manifold as the underlying structure
of the model. Only manifolds of the dimension four have an infinity
of possible non-diffeomorphic smoothness structures (see for instance
\cite{Gom:85,Tau:87,FiSt:96}). For all other dimensions, there are
only finite many smoothness structures \cite{KerMil:63,KirSie:77}.
This richness of the smoothness structures in four dimensions should
have a physically meaning. There is a growing literature discussing
the influence of exotic smoothness on physics. It is a common believe
that exotic smoothness is a main contribution to the path integral
of quantum gravity which was confirmed in a special case by one of
the authors \cite{Ass2010}. The topic started with the paper of Brans
and Randall \cite{BraRan:93} and later by Brans \cite{Bra:94a,Bra:94b}
leading to the guess that exotic smoothness can be a source of non-standard
solutions of Einsteins equation (Brans conjecture). One of the authors
published an article \cite{Ass:96} to show the influence of the differential
structure to GRT for compact manifolds of simple type. S{\l}adkowski
showed in \cite{Sladkowski2001} that the exotic ${\mathbb{R}}^{4}$
can act as the source of the gravitational field. Thus exotic smoothness
is able to represent a source of a gravitational field which cannot
be distinguished from a usual source by an external observer. Furthermore,
these sources are localized in the 4-manifold, i.e. one can construct
a non-diffeomorphic smoothness structure from a given one by a modification
of a 4-dimensional submanifold. As far as we know, sources of a gravitational
field are any kind of matter (baryonic, radiation or dark). From all
this it seems natural to relate matter with exotic smoothness. We
will support this conjecture by showing that a 4-manifold admitting
a Ricci-flat metric (in standard smoothness structure) changes to
a 4-manifold with non-Ricci-flat metric in all exotic smoothness structures
(see the discussion in subsection \ref{sub:Motivation}). Thus \emph{if
one starts with a vacuum solution of Einsteins equation then exotic
smoothness modifies this solution to a non-vacuum solution, i.e. the
sources are determined by the exotic smoothness structure}. In the
following we will study this relation more carefully by showing how
exotic smoothness is able to generate the known action terms.

\subsection{Outline of the paper}

At first we have to note that there are many examples of exotic, \emph{non-compact}
4-manifolds which are hard to describe at the present state of knowledge.
Thus we will restrict ourself to the class of \emph{compact} 4-manifolds
where one has powerful invariants like Seiberg-Witten invariants \cite{Akb:96,Wit:94SW}
or Donaldson polynomials \cite{Don:90,DonKro:90} to distinguish between
different smoothness structures. A recent result \cite{yasui2011}
can be used to produce a non-compact exotic 4-manifold (distinguished
by the Seiberg-Witten invariants) which embeds in the compact 4-manifold
considered. So, if we have a compact 4-manifold $M$ then the non-compact
4-manifold $M\setminus B^{4}$ (with the open 4-ball $B^{4}=D^{4}\setminus\partial D^{4}$)
embeds obviously in $M$. Especially if $M$ is exotic then $M\setminus B^{4}$
is exotic too. Additionally the non-compact 4-manifold $M\setminus B^{4}$
admits a Lorentz metric.

Secondly, we will not discuss the definition of the path integral
and all problems connected with renormalization, definiteness etc.

Third, we use the knot surgery of Fintushel and Stern \cite{FiSt:96}
to construct the exotic smoothness structure. This approach assumes
a special class of 4-manifolds ({}``complicated-enough'') which
contains the class of elliptic surfaces among them the important K3
surface.

This leading us to the assumptions of the 4-manifold representing
the spacetime of the model:
\begin{enumerate}
\item The compact and simple-connected 4-manifold under consideration is
an elliptic surface $E(n)$,especially $E(2)$ which is the \emph{K3
surface}.
\item The non-compact 4-manifold $E(n)\setminus B^{4}$ admits a Lorentz
metric and represents the spacetime of the model.
\item The exotic smoothness structures of the 4-manifold $E(n)$ are constructed
by Fintushel-Stern knot surgery. It follows that this surgery is also
producing an exotic smoothness structure on $E(n)\setminus B^{4}$.
\end{enumerate}
In section \ref{sec:Matter-as-exotic-space} we will describe the
physical model and motivate it. As shown by LeBrun \cite{Lebrun96,Lebrun98}
using Seiberg-Witten theory, a 4-manifold as above admitting an Einstein
metric in the standard smooth structure fails to admit an Einstein
metric in some exotic smooth structure. Therefore a spacetime without
matter can be changed to a spacetime with source terms by changing
the smoothness structure. Together with the confirmed Brans conjecture,
that exotic smoothness is a source of gravity, we have a good motivation
for our main hypothesis: \emph{matter as represented by fields is
emerged from exotic smoothness}. With this motivation in mind, we
start with a 4-manifold $M=E(2)$ in standard structure and apply
the Fintushel-Stern knot surgery to get the exotic 4-manifold $M_{K}$
and consider the Einstein-Hilbert action on $M_{K}$ agreeing with
the Einstein-Hilbert action on $M$ outside of some compact submanifold.
Then we will discuss in section \ref{sec:Geometrical-interpretation}
a decomposition of the 4-manifold $M_{K}$ which induces by using
the diffeomorphism invariance a decomposition of the action. In this
decomposition the knotted torus of the knot surgery corresponds to
the Dirac action via its Weierstrass (or spinor) representation, described
in section \ref{sec:Dirac-action}. For more complicated knots like
satellite knots or sums of knots we obtain a splitting of the knot
complement into simpler pieces clued together along connecting tubes.
In section \ref{sec:Gauge-field-action} we show that this connecting
tubes represent the gauge fields and in section \ref{sec:Gauge-group}
we will speculate about the derivation of the gauge group using the
classification of the connecting tubes. A short discussion of the
results in the last section completes the paper.

\section{Lorentz metric and global hyperbolicity\label{sec:Lorentz-metric-global-hyp}}

Before we start the investigation of the proposed model, we will discuss
some more general physical implications. Firstly we consider the existence
of a Lorentz metric, i.e. a 4-manifold $M$ (the spacetime) admits
a Lorentz metric if (and only if) there is a non-vanishing vector
field. In case of a compact 4-manifold $M$ we can use the Poincare-Hopf
theorem to state: a compact 4-manifold admits a Lorentz metric if
the Euler characteristic vanishes $\chi(M)=0$. But in a compact 4-manifold
there are closed time-like curves (CTC) contradicting the causality
or more exactly: the chronology violating set of a compact 4-manifold
is non-empty (Proposition 6.4.2 in \cite{HawEll:94}). Non-compact
4-manifolds $M$ admit always a Lorentz metric and a special class
of these 4-manifolds have an empty chronology violating set. If $\mathcal{S}$
is an acausal hypersurface in $M$ (i.e., a topological hypersurface
of $M$ such that no pair of points of $M$ can be connected by means
of a causal curve), then $D^{+}(\mathcal{S})$ is the future Cauchy
development (or domain of dependence) of $\mathcal{S}$, i.e. the
set of all points $p$ of $M$ such that any past-inextensible causal
curve through $p$ intersects $\mathcal{S}$. Similarly $D^{-}(\mathcal{S})$
is the past Cauchy development of $\mathcal{S}$. If there are no
closed causal curves, then $\mathcal{S}$ is a Cauchy surface if $D^{+}(\mathcal{S})\cup\mathcal{S}\cup D^{-}(\mathcal{S})=M$.
As shown in \cite{BernalSanchez2003}, the existence of a Cauchy surface
implies that $M$ is diffeomorphic to $\mathcal{S}\times\mathbb{R}$
. 

This strong result is also connected with the concept of global hyperbolicity.
A spacetime manifold $M$ without boundary is said to be \emph{globally
hyperbolic} if the following two conditions hold:
\begin{enumerate}
\item \emph{Absence of naked singularities}: For every pair of points $p$
and $q$ in $M$, the space of all points that can be both reached
from $p$ along a past-oriented causal curve and reached from $q$
along a future-oriented causal curve is compact.
\item \emph{Chronology}: No closed causal curves exist (or ''Causality'' holds on $M$).
\end{enumerate}
Usually condition 2 above is replaced by the more technical condition
''Strong causality holds on $M$'' but as
shown in \cite{BernalSanchez2007} instead of ''strong causality'' one can write simply the condition ''causality''
(and strong causality will hold under causality plus condition 1 above). 

Then together with the diffeomorphism between $M$ and $\mathcal{S}\times\mathbb{R}$
we can conclude that all (non-compact) 4-manifolds $\mathcal{S}\times\mathbb{R}$
are the only 4-manifolds which admit a globally hyperbolic Lorentz
metric \cite{BernalSanchez2003}. The existence of a Cauchy surface
$\mathcal{S}$ implies global hyperbolicity of the spacetime and its
unique representation by $\mathcal{S}\times\mathbb{R}$ (up to diffeomorphism).
But as shown in \cite{BernalSanchez2005}, also the metric is determined
(up to isometry) by global hyperbolicity.

\begin{theorem}\label{thm:global-hyp}

If a spacetime $(M,g)$ is globally hyperbolic, then it is isometric
to $(\mathbb{R}\times\mathcal{S},-f\cdot d\tau^{2}+g_{\tau})$
with a smooth positive function $f:\mathbb{R}\to\mathbb{R}$ and a
smooth family of Riemannian metrics $g_{\tau}$ on $\mathcal{S}$
varying with $\tau$. Moreover, each $\left\{ t\right\} \times\mathcal{S}$
is a Cauchy slice.

\end{theorem} Furthermore in \cite{BernalSanchez2006} it was shown:
\begin{itemize}
\item If a compact spacelike submanifold with boundary of a globally hyperbolic
spacetime is acausal then it can be extended to a full Cauchy spacelike
hypersurface $\mathcal{S}$ of $M$, and
\item for any Cauchy spacelike hypersurface $\mathcal{S}$ there exists
a function as in Th. \ref{thm:global-hyp} such that $\mathcal{S}$
is one of the levels $\tau=constant$.
\end{itemize}
But what are the implications of global hyperbolicity in the exotic
case? At first, the existence of a Lorentz metric is a purely topological
condition which will be fulfilled by all non-compact 4-manifolds independent
of the smoothness structure. But by considering global hyperbolicity
the picture changes. An exotic spacetime $M=(\mathcal{S}\times\mathbb{R})_{exotic}$
homeomorphic to $\mathcal{S}\times\mathbb{R}$ is \emph{not diffeomorphic}
to $\mathcal{S}\times\mathbb{R}$. The Cauchy surface $\mathcal{S}$
is a 3-manifold with an unique smoothness structure (up to diffeomorphisms)
-- the standard structure. So, the smooth product $\mathcal{S}\times\mathbb{R}$
must be admit the standard smoothness structure. But the diffeomorphism
\cite{BernalSanchez2003} between $M$ and $\mathcal{S}\times\mathbb{R}$
is necessary for global hyperbolicity. Therefore an \emph{exotic $(\mathcal{S}\times\mathbb{R})_{exotic}$
is never globally hyperbolic but admits a Lorentz metric}. Generally
we have an exotic \emph{$(\mathcal{S}\times\mathbb{R})_{exotic}$}
with a Lorentz metric such that the projection $(\mathcal{S}\times\mathbb{R})_{exotic}\to\mathbb{R}$
is a time-function (that is, a continuous function which is strictly
increasing on future directed causal curves). But then the exotic
\emph{$(\mathcal{S}\times\mathbb{R})_{exotic}$} has no closed causal
curves and must contain naked singularities%
\footnote{Any non-compact manifold $M$ admits stably causal metrics (that is,
those with a time function). So, if $M$ is not diffeomorphic to some
product$\mathcal{S}\times\mathbb{R}$, all these (causally well behaved)
metrics must contain naked singularities. We thank M. S\'anchez for
the explanation of this result.%
}. 

\begin{figure}
\includegraphics[angle=90,scale=0.25]{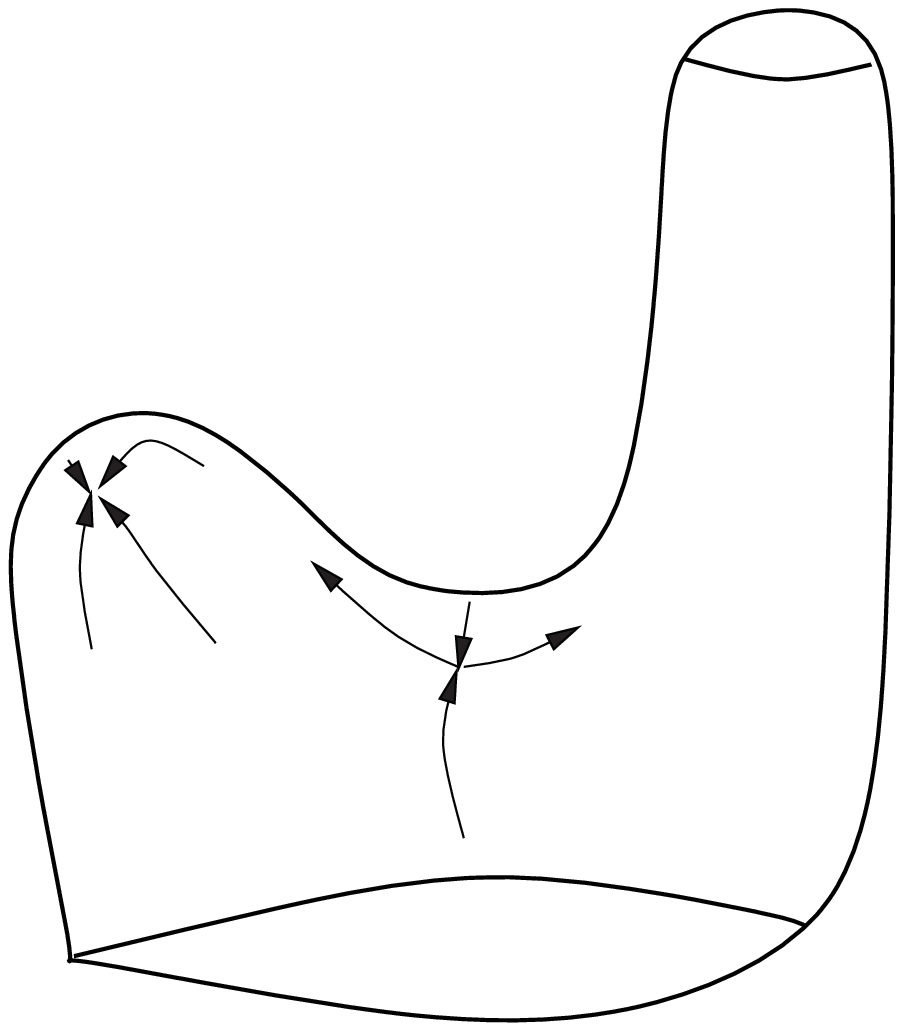}

\caption{Two naked singularities.\label{fig:toy-naked-singularity}}

\end{figure}

\begin{figure}
\includegraphics{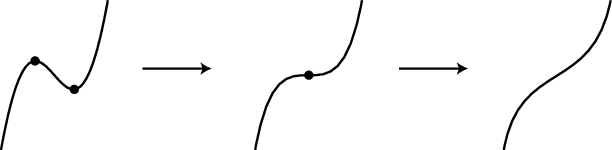}

\caption{Killing a 0- and a 1-handle.\label{fig:killling-handles}}

\end{figure}

With this result in mind, one should ask for the physical interpretation
of naked singularities. To visualize the problem, we consider the
following toy model: a non-trivial surface (see Fig. \ref{fig:toy-naked-singularity})
connecting two circles which can be deformed to the usual cylinder.
 This example can be described by the concept of a cobordism. A cobordism
$(W,M_{1},M_{2})$ between two $n-$manifolds $M_{1},M_{2}$ is a
$(n+1)-$manifold $W$ with $\partial W=M_{1}\sqcup M_{2}$ (ignoring
the orientation). Then there exists a smooth function $f:W\to[0,1]$
with isolated critical points (vanishing first derivative) such that
$f^{-1}(0)=M_{1},\, f^{-1}(1)=M_{2}$. By general position arguments,
one can assume that all critical points of $f$ occur in the interior
of $W$. In this setting $f$ is called a Morse function on a cobordism.
For every critical point of $f$ (vanishing first derivative) one
adds a so-called $k-$handle $D^{k}\times D^{n-k}$. In our example
in Fig. \ref{fig:toy-naked-singularity}, we add a 2-handle $D^{2}\times D^{0}$
(the maximum) and a 1-handle $D^{1}\times D^{1}$ (the saddle). But
obviously this cobordism is diffeomorphic to the trivial cobordism
$S^{1}\times[0,1]$ (the two boundary components are diffeomorphic
to each other). Therefore the 2-/1-handle pair is ``killed\textasciiacute{}\textasciiacute{}
in this case. The 2-handle and the 1-handle differ in one direction
whereas the Morse function has a maximum for the 2-handle and a minimum
for the 1-handle. The left graph of Fig. \ref{fig:killling-handles}
visualizes this fact. %
 Furthermore the sequence of graphs from the left to right represents
the process to ''kill'' the handle pair. 

In our example of an exotic $(\mathcal{S}\times\mathbb{R})_{exotic}$,
we consider a (non-compact) cobordism between homeomorphic boundary
components (the two Cauchy surfaces at infinity, i.e. ''$\mathcal{S}\times\left\{ -\infty\right\} $
and $\mathcal{S}\times\left\{ +\infty\right\} $''). By a result of
\cite{Mil:65}, a cobordism of this kind (a so-called h-cobordism)
contains only handles of complement dimension in the interior. But
these handles can be killed: the details of the construction can be
found in \cite{Mil:65}. Here we will give only some general remarks.
Any $0-/1-$handle pair as well any $n-/(n+1)-$handle pair (remember
the h-cobordism is $n+1$-dimensional) can be killed by a general
procedure. The killing of a $k-/(k+1)-$handle pair depends on a special
procedure, the Whitney trick. For 4- and 5-dimensional h-cobordisms
(between 3- and 4-manifolds, respectively) we cannot use the Whitney
trick. This failure lies at the heart of the problem to classify 3-
and 4-manifolds (see \cite{GomSti:1999}).

In case of the spacetime we are interested in a 4-dimensional h-cobordism
between 3-manifolds. Here we can kill the $0-/1-$and the $3-/4-$handle
pair of the h-cobordism. As a result we get extra $1-/2-$handle and
$2-/3-$handle pairs. If the Whitney trick works in this case, we
can also kill these pairs of handles. But it is known that the Whitney
trick only works topologically \cite{Fre:82}. The exotic manifolds
$(\mathcal{S}\times\mathbb{R})_{exotic}$ (as non-compact examples)
are counterexamples that the (infinite) pairs of handles never cancel
each other. Lets start with $1-/2-$handle pair. The critical point
of the Morse function with index $1$ (the Morse function has a minimum
in one directions (saddle point)) corresponds to the 1-handle. Then
the Morse function of index 2 corresponds to the 2-handle. Each pair
of handles is connected to each other, i.e. the direction representing
the minimum of a 1-handle is connected with one direction representing
the maximum of the 2-handle. This connection between the 1- and the
2-handle is non-trivial, i.e. there are extra intersection points
between the (attaching) sphere of the 1-handle and the (belt) sphere
of the 2-handle (see \cite{GomSti:1999} Prop. 4.2.9). The $2-/3-$handle
pair is dual to the $1-/2-$handle pair, i.e. the number of $2-/3-$handle
pairs is equal to the number of $1-/2-$handle pairs. Each handle
of the pair represents the neighborhood of a naked singularity. Then
these singularities appear only pairwise. The Morse vector field (gradient
of the Morse function) vanishes but we get a saddle having negative
curvature (\emph{no curvature singularity} or \emph{quasiregular singularity,}
also called \emph{locally extendible singularity} \cite{Ellis1977}).
There is growing evidence for the appearance of this singularity without
violating causality (causal continuity, see \cite{Dowker1997}). Thus
it is very probable that the property of exotic smoothness requiring
the appearance of naked singularities in form of non-canceling 2-handle
pairs can be interpreted in a consistent physically way: In the following
sections we will merge exotic smoothness with fermion and bosons obtaining
the consequence that exotic smoothness implies naked singularities
as well as physically particles. We do not think this is an accident
-- we conjecture that the naked singularities can be seen as the geometrically
consequence of the physically particles.

In this paper we propose a model where an open 4-ball $B^{4}$ ($B^{4}=D^{4}\setminus\partial D^{4}$)
is removed from the compact 4-manifold $M$ to obtain the open 4-manifold
$M\setminus B^{4}$. This 4-manifold admits a Lorentz metric but contains
also naked singularities. In the following we discuss the case of
a compact 4-manifolds admitting a Ricci-flat metric, the K3 surface.
The choice has a conceptional background: If we interpret matter as
geometrical objects then there is no need for a non-geometrical energy-momentum
tensor. Therefore we have to consider the source-free Einstein equation
implying Ricci-flatness. But there are only two compact 4-manifolds
with Ricci-flat metric, the 4-torus and the K3 surface. This result
is only true for a Riemannian metric but we can introduce a Lorentz
metric for the K3 surface and the 4-torus by removing an open 4-ball
(see above). The 4-torus is a flat manifold, i.e. the curvature vanishes.
From the physical point of view, it is useless to choose this manifold.
Therefore we will concentrate on the K3 surface (minus an open 4-ball)
which is a so-called elliptic surface.

\section{Preliminaries: Elliptic surfaces and exotic smoothness}

In this section we will give some information about elliptic surfaces
and its construction. First we will discuss smoothness in general
and give a short overview of the construction of elliptic surfaces
and a special class of elliptic surfaces denoted by $E(n)$ in the
literature. Then we present the construction of exotic $E(n)$ by
using the knot surgery of Fintushel and Stern.

\subsection{Smoothness on manifolds}

From the mathematical point of view, the spacetime is a smooth 4-manifold
endowed with a (smooth) metric as basic variable for general relativity.
In the previous section we discuss the existence question for Lorentz
structure and causality problems (see Hawking and Ellis \cite{HawEll:94})
giving further restrictions on the 4-manifold: causality implies non-compactness,
Lorentz structure needs a non-vanishing normal vector field. The appropriate
notation is the global hyperbolic 4-manifold $M$ having a Cauchy
surface $\mathcal{S}$ so that $M=\mathcal{S}\times\mathbb{R}$. 

All these restrictions on the representation of spacetime by the manifold
concept are clearly motivated by physical questions. Among the properties
there is one distinguished element: the smoothness. Usually one starts
with a topological 4-manifold $M$ and introduces structures on them.
Then one has the following ladder of possible structures:\begin{eqnarray*}
\mbox{Topology}\to & \mbox{\mbox{piecewise-linear(PL)}}\to & \mbox{Smoothness}\to\\
\qquad\to & \mbox{bundles, Lorentz, Spin etc.}\to & \mbox{metric, geometry,...}\end{eqnarray*}
We do not want to discuss the first transition, i.e. the existence
of a triangulation on a topological manifold. But we remark that the
existence of a PL structure implies uniquely a smoothness structure
in all dimensions smaller than 7 \cite{KirSie:77}. Here we have to
consider the following steps to define a spacetime:
\begin{enumerate}
\item Fix a topology for the spacetime $M$.
\item Fix a smoothness structure, i.e. a maximal differentiable atlas $\mathcal{A}$.
\item Fix a smooth metric or get one by solving the Einstein equation.
\end{enumerate}
The choice of a topology never fixes the spacetime uniquely, i.e.
there are two spacetimes with the same topology which are not diffeomorphic.
The following basic facts should the reader keep in mind for any $n-$dimensional
manifold $M^{n}$:
\begin{enumerate}
\item The maximal differentiable atlas $\mathcal{A}$ of $M^{n}$ is the
smoothness structure.
\item To determine a smoothness structure it suffices to give a single maximal
differentiable atlas. Thus $\mathbb{R}^{n}$ has an unique smoothness
structure containing the identity map of $\mathbb{R}^{n}$ (\emph{standard
smoothness structure}, $(\mathbb{R}^{n},id_{\mathbb{R}^{n}})$ is
the atlas). 
\item It is difficult to define the \emph{standard smoothness structure}
on a general 4-manifold $M$. One way to get around this difficulty
is the usage of the instability of all exotic smoothness structures
in dimension 4. Stable smoothness structures are able to extend from
a smoothness structure on $M$ to $M\times\mathbb{R}^{k}$ (see \cite{KirSie:77}
for the notation of a stable CAT structure). The classification theory
of smoothness structure \cite{Mun:60,KirSie:77} for all manifolds
of dimension greater than $5$ implies (together with a result of
Quinn \cite{Qui:82} about the vanishing of $\pi_{4}(TOP/O)=0$) that
the smoothness structure of $M\times\mathbb{R}^{k}$ is unique for
all $k>0$ (up to diffeomorphisms). Here we have to assume that the
Kirby-Siebenmann invariant vanishes. We call this smoothness structure
the standard smoothness structure of $M\times\mathbb{R}^{k}$. Then
one can extend this smoothness structure to $M$ by restriction. All
other possible smoothness structures non-diffeomorphic to the standard
smoothness structure are called \emph{exotic} smoothness structures.
\item The existence of a smoothness structure is \emph{necessary} to introduce
Riemannian or Lorentzian structures on $M$, but the smoothness structure
do not further restrict the Lorentz structure.
\end{enumerate}
We want to close this subsection with a general remark: the number
of non-diffeomorphic smoothness structures is finite for all dimensions
$n\not=4$ \cite{KirSie:77}. In dimension four there are many examples
of compact 4-manifolds with infinite finite and many examples of non-compact
4-manifolds with uncountable infinite many non-diffeomorphic smoothness
structures.

\subsection{Elliptic surfaces \label{sec:log} }

A \emph{complex surfaces} $S$ is a 2-dimensional complex manifold
which is compact and connected. A special complex surface is the \emph{elliptic
surface}, i.e. a complex surface $S$ together with a map $\pi:S\to C$
($C$ complex curve, i.e. Riemannian surface), so that for nearly
every point $p\in S$ the reversed map $F=\pi^{-1}(p)$ is an elliptic
curve, i.e. a torus%
\footnote{We denote this map $\pi$ as elliptic fibration. Thus every complex
surface which is equipped with a elliptic fibration is an elliptic
surface.%
}. Now we will construct the special class of elliptic surfaces $E(n)$.

The first step is the construction of $E(1)$ by the unfolding of
singularities for two cubic polynomials intersecting each other. The
resulting manifold $E(1)$ is the manifold ${\mathbb{C}P}^{2}\#9\overline{\mathbb{C}P}^{2}$
but equipped with an elliptic fibration. Then we use the method of
fiber sum to produce the surfaces $E(n)$ for every number $n\in\mathbb{N}$.
For that purpose we cut out a neighborhood $N(F)$ of one fiber $\pi^{-1}(p)$
of $E(1)$. Now we sew together two copies of $E(1)\setminus N(F)$
along the boundary of $N(F)$ to get $E(2)$, i.e. we define the fiber
sum $E(2)=E(1)\#_{f}E(1)$. Especially we note that $E(2)$ is also
known as \emph{K3-surface} widely used in physics. In general we get
the recursive definition $E(n)=E(n-1)\#_{f}E(1)$. The details of
the construction can be found in the paper \cite{Gom:91.2}.

\subsection{Knot surgery and exotic elliptic surface\label{sub:Knot-surgery-elliptic-surface}}

The main technique to construct an exotic elliptic surface was introduced
by Fintushel and Stern \cite{FiSt:96}, called \emph{knot surgery}.
In short, given a simple-connected, compact 4-manifold $M$ with an
embedded torus $T^{2}$ (having special properties, see below), cut
out $M\setminus N(T^{2})$ a neighborhood $N(T^{2})=D^{2}\times T^{2}$
of the torus and glue in $S^{1}\times\left(S^{3}\setminus N(K)\right)$.
The 3-manifold $S^{3}\setminus N(K)$ is called the \emph{knot complement}
of $K$ (see appendix \ref{sec:Knot-complement})%
\footnote{Alternatively, the knot complement is the 3-manifold obtained by performing
0-framed surgery on the knot.%
}. This surgery construction depends only on the gluing operation of
the knot $K$, i.e. an embedding of the circle $S^{1}$ into $\mathbb{R}^{3}$
or $S^{3}$, and one obtains the new 4-manifold\[
M_{K}=(M\setminus N(T))\cup_{T^{3}}(S^{1}\times(S^{3}\setminus N(K)))\]
from a given 4-manifold $M$ by gluing $M\setminus N(T^{2})$ and
$S^{1}\times\left(S^{3}\setminus N(K)\right)$ along the common boundary,
the 3-torus $T^{3}$. The remarkable result of Fintushel and Stern
\cite{FiSt:96} is, that a gluing with a non-trivial knot $K$ changes
$M$ non-diffeomorphic to $M_{K}$. We remark that the construction
can be easily generalized to links, i.e. the embedding of the disjoint
union of circles $S^{1}\sqcup\cdots\sqcup S^{1}$ into $\mathbb{R}^{3}$
or $S^{3}$. The reader not interested in the details of the construction
can now jump to the next section.

The precise definition can be given in the following way for elliptic
surface: Let $\pi\colon S\to C$ be an elliptic surface and $\pi^{-1}(t)=F$
a smooth fiber $(t\in C)$. As usual, $N(F)$ denotes a neighborhood
of the regular fiber $F$ in $S$ (which is diffeomorphic to $D^{2}\times T^{2}$).
Deleting $N(F)$ from $S$ to get a manifold $S\setminus N(F)$ with
boundary $\partial(S\setminus N(F))=\partial(D^{2}\times F)=S^{1}\times F=T^{3}$,
the 3-torus. Then we take the 4-manifold $S^{1}\times(S^{3}\setminus N(K))$,
$K$ a knot, with boundary $\partial(S^{1}\times(S^{3}\setminus N(K)))=T^{3}$
and regluing it along the common boundary $T^{3}$. The resulting
4-manifold\[
S_{K}=(S\setminus N(F))\cup_{T^{3}}(S^{1}\times(S^{3}\setminus N(K)))\]
is obtained from $S$ by knot surgery using the knot $K$. The regular
fiber $F$ in the elliptic surface $S$ has two properties which are
essential for the whole construction:
\begin{enumerate}
\item In a larger neighborhood $N_{c}(F)$ of the regular fiber $F$ there
is a cups fiber $c$, i.e. an embedded 2-sphere of self-intersection
$0$ with a single, non-locally flat point whose neighborhood is the
cone over the right-hand trefoil knot (so-called c-embedded torus).
\item The complement $S\setminus F$ of the regular fiber is simple-connected
$\pi_{1}(S\setminus F)=1$.
\end{enumerate}
Then as shown in \cite{FiSt:96}, $S_{K}$ is not diffeomorphic to
$S$. The whole procedure can be generalized to any 4-manifold allowing
an embedding of a torus of self-intersection $0$ in a neighborhood
of a cusp. 

Before we proceed with the physical interpretation, we will discuss
the question when two exotic $S_{K}$ and $S_{K'}$ for two knots
$K,\, K'$ are diffeomorphic to each other. Currently there are two
invariants to distinguish non-diffeomorphic smoothness structures:
Donaldson polynomials and Seiberg-Witten invariants. Fintushel and
Stern \cite{FiSt:96} calculated the Seiberg-Witten invariants for
$S_{K}$ and $S_{K'}$ to show that $S_{K}$ differs from $S_{K'}$
if the Alexander polynomials (a knot invariant, see \cite{Rol:76})
of the two knots differ. Unfortunately this invariant is not a complete
classifying invariant for knots. Thus we cannot say anything about
$S_{K}$ and $S_{K'}$ for two knots with the same Alexander polynomial.
But Fintushel and Stern \cite{FinSter:1999,FinSter:2002} constructed
counterexamples of two knots $K,\, K'$ with the same Alexander polynomial
but with different $S_{K}$ and $S_{K'}$. Furthermore Akbulut \cite{Akb:99}
showed that the knot $K$ and its mirror $\bar{K}$ induce diffeomorphic
4-manifold $S_{K}=S_{\bar{K}}$. 

In a recent paper \cite{yasui2011}, the non-compact case was also
discussed. Consider the non-compact elliptic surface $S_{K}\setminus B^{4}$
which embeds in $S_{K}$. By Theorem 1.1 in \cite{yasui2011}: if
the Seiberg-Witten invariant of $S_{K}$ do not vanish (which is true
for a knot $K$ with non-trivial Alexander polynomial) then $S_{K}\setminus B^{4}$
admits infinitely many distinct exotic smooth structures.

\section{Matter from exotic spacetime\label{sec:Matter-as-exotic-space}}

\subsection{Motivation\label{sub:Motivation}}

In this section we will motivate our hypothesis: matter emerges from
exotic spacetime. We start with the Brans conjecture: exotic smoothness
induces additional sources of gravity. One of the authors proved this
conjecture for compact manifolds \cite{Ass:96} generating source
terms in Einstein field equation. Then Sladkowski \cite{Sladkowski2001}
showed the conjecture for the exotic $\mathbb{R}^{4}$, i.e. the exoticness
implies non-flat solutions of the Einstein field equation. Now we
will discuss properties of matter and its realization by exotic smoothness:
\begin{enumerate}
\item Locality: The smoothness structure is by definition a global structure
-- the maximal differentiable atlas. But in knot surgery, a local
modification changes the smoothness structure. 
\item Infinite trajectories: As discussed above, we consider a non-compact
4-manifold as spacetime. Therefore, we have trajectories which can
be extended to infinity.
\item Stability: Every knot surgery enforces a modification of the smoothness
structure to a non-diffeomorphic one in relation to the smoothness
structure at the starting point. Therefore it seems that exotic smoothness
is a relative phenomenon which can be gauged away -- but this is wrong.
The necessary transformation for changing the smoothness is a non-diffeomorphism,
i.e. one changes the whole physical system by using this transformation.
Secondly, we will show in this paper that every exotic smoothness
structure generates matter. Then the standard smoothness structure
represents the spacetime without matter.
\end{enumerate}
In subsection \ref{sub:Knot-surgery-elliptic-surface} we discuss
the smoothness structures for a compact, simple-connected, smooth
4-manifold by using the Seiberg-Witten invariant \cite{Wit:94SW,Akb:96}.
It is known that an exotic smoothness structure implies a non-trivial
Seiberg-Witten invariant, i.e. this invariant vanishes for the standard
structure. While the invariant is not complete (two different exotic
smoothness structures can have the same invariant) it is able to distinguish
between the standard smoothness structure and any exotic one. 

In this paper we will consider the K3 surface $E(2)$ and choose its
non-compact version $E(2)\setminus B^{4}$ as spacetime model. The
choice of the K3 surface is not arbitrary: the K3 surface is the only
compact, simple-connected, closed 4-manifold with Ricci-flat metric
\cite{Besse1987} (see also the section \ref{sec:Lorentz-metric-global-hyp}).
This Ricci-flat metric is a Riemannian metric but this is no problem
because the K3 surface itself do not admit a Lorentz metric at all.
Hence one has to choose as a model of the spacetime the non-compact
version of the K3 surface $E(2)\setminus B^{4}$ admitting a Lorentz
metric. In the usual view, matter is represented by a source-term
of the Einstein equation. In the geometrical model proposed here it
should be a expression of exotic smoothness only without the help
of conventional (non-geometrical) source-terms. Thus to investigate
the effect of exoticness only we have to start with a plain spacetime
in standard smoothness structure and with vacuum metric, i.e. an Einstein
equation without any conventional matter-terms. Then we can evaluate
the effect of the exoticness by switching this vacuum spacetime to
a exotic smoothness structure. For this propose our starting spacetime
has to admit a vacuum solution, i.e. has to be Ricci-flat. As mention
above, the K3 surface is the only compact, simple-connected, closed
4-manifold with Ricci-flat metric. This motivates the choice.

\subsection{The model}

In this section we will discuss the additional contribution to the
Einstein-Hilbert action functional coming from exotic smoothness generated
by knot surgery. Our model starts with an elliptic surface, the K3
surface $E(2)$ motivated in the previous subsection. We choose the
non-compact 4-manifold $M=E(2)\setminus B^{4}$ as spacetime, admitting
a Ricci-flat metric $g$. The work in \cite{yasui2011} relates the
smoothness properties of $E(2)$ to $M$. Especially a knot surgery
on $E(2)$ produces an exotic smoothness structure (if the knot has
non-trivial Alexander polynomial) as well on the spacetime $M$. Here
we consider the Einstein-Hilbert action\begin{equation}
S_{EH}(g)=\intop_{M}R\sqrt{g}\: d^{4}x\label{eq:EH-action}\end{equation}
with the Ricci-flat metric $g$ as solution of the vacuum field equations
and study the effect of switching the smooth structure by a knot surgery.
This procedure touches only a submanifold $N(T^{2})\subset M$ and
thus we consider a decomposition of the 4-manifold\[
M=(M\setminus N(T^{2}))\cup_{T^{3}}N(T^{2})\]
with $N(T^{2})=D^{2}\times T^{2}$ leading to a sum in the action
\begin{equation}
S_{EH}(M)=\intop_{M\setminus N(T^{2})}R\sqrt{g}\: d^{4}x+\intop_{N(T^{2})}R\sqrt{g}\: d^{4}x\quad.\label{eq:relation-action-M}\end{equation}
Because of diffeomorphism invariance of the Einstein-Hilbert action,
this decomposition do not depend on the concrete realization with
respect to any coordinate system. Now we switch to a new smoothness
structure on $M$ by using a knot $K:S^{1}\to S^{3}$ \[
M_{K}=(M\setminus N(T^{2}))\cup_{T^{3}}(S^{1}\times(S^{3}\setminus N(K)))\]
i.e. we exchange $N(T^{2})$ by $S^{1}\times(S^{3}\setminus N(K))$
and call the resulting exotic 4-manifold $M_{K}$. The 4-manifold
$M\setminus N(T^{2})$ with boundary a 3-torus $T^{3}$ appears in
both 4-manifolds $M$ and $M_{K}$. Thus we can fix its action\[
S_{EH}(M\setminus N(T^{2}))=\intop_{M\setminus N(T^{2})}R\sqrt{g}\, d^{4}x\]
by using a fixed metric $g$ in the interior $int(M\setminus N(T^{2}))$.
Furthermore because of the identy gluing map $T^{3}=\partial(M\setminus N(T^{2}))\to\partial N(T^{2})=T^{3}$
the boundary terms for $M\setminus N(T^{2})$ and $N(T^{2})$ are
equal but have a different orientation and therefore cancel each other.
The decomposition of $M_{K}$ gives for the action\begin{equation}
S_{EH}(M_{K})=S_{EH}(M\setminus N(T^{2}))+\intop_{S^{1}\times(S^{3}\setminus N(K))}R_{K}\sqrt{g_{K}}\, d^{4}x\label{eq:relation-action-1}\end{equation}
with a metric $g_{K}$ and scalar curvature $R_{K}$ of the 4-manifold
$S^{1}\times(S^{3}\setminus N(K))$. To determine the first term in
(\ref{eq:relation-action-1}) by using (\ref{eq:relation-action-M})
we consider the integral\begin{equation}
S_{EH}(N(T^{2}))=\intop_{N(T^{2})=D^{2}\times T^{2}}R\sqrt{g}\, d^{4}x\label{eq:flat-torus}\end{equation}
over $N(T^{2})=D^{2}\times T^{2}$. By definition $N(T^{2})$ is a
submanifold of $M$. Then the torus $T^{2}$ in $N(T^{2})=T^{2}\times D^{2}$
can be seen as an embedded (unknotted) 2-manifold $T^{2}\hookrightarrow N(T^{2})$.
We choose a product metric for $N(T^{2})$ to calculate the integral
(\ref{eq:flat-torus}) where the metric of the embedded torus is induced
by restriction. Then by using the results in \cite{KuiperMeeks1987}
we obtain\[
S_{EH}((N(T^{2}))=vol(D^{2})\cdot\intop_{T^{2}\subset N(T^{2})}K_{T^{2}}d\sigma_{T^{2}}\]
the total curvature of the embedded torus ($K_{T^{2}}$ is the Gaussian
curvature of the embedded torus, $d\sigma_{T^{2}}$ is the area element).
Let $\mathbf{K}$ be the extrinsic curvature of the embedded torus
$T^{2}\subset N(T^{2})$ then we obtain alternatively\[
S_{EH}((N(T^{2}))=vol(D^{2})\cdot\intop_{T^{2}}((tr\mathbf{K})^{2}-tr\mathbf{K}^{2})\, d\sigma_{T^{2}}\]
The embedded torus is unknotted and the greatest lower bound for the
total curvature of this torus is $8\pi$ (see \cite{KuiperMeeks1987}).
Thus, the integral (\ref{eq:flat-torus}) has a constant value%
\footnote{We thank the referee for pointing out an error in our previous argumentation.%
} \[
S_{EH}(N(T^{2}))=\lambda\cdot vol(D^{2})=const.\]
 and by using (\ref{eq:relation-action-M}) we have \[
S_{EH}(M\setminus N(T^{2}))=S_{EH}(M)-\lambda\cdot vol(D^{2})\quad.\]
Using this and (\ref{eq:relation-action-1}), we obtain the relation\begin{equation}
S_{EH}(M_{K})=S_{EH}(M)+\intop_{S^{1}\times(S^{3}\setminus N(K))}R_{K}\sqrt{g_{K}}\, d^{4}x-\lambda\cdot vol(D^{2})\label{eq:relation-action-knot-surgery}\end{equation}
between the Einstein-Hilbert action on $M$ and $M_{K}$ showing that
the \emph{contribution of the exotic smoothness structure to the Einstein-Hilbert
action} is given by the action integral over $S^{1}\times(S^{3}\setminus N(K))$
and a constant $\lambda\cdot vol(D^{2})$. We evaluate the exotic
action \begin{equation}
S_{exotic}\doteq\intop_{S^{1}\times(S^{3}\setminus N(K))}R_{K}\sqrt{g_{K}}\, d^{4}x-\lambda\cdot vol(D^{2})\label{eq:EH-action-knot-complement}\end{equation}
by using a product metric $g_{K}$\begin{equation}
ds^{2}=d\theta^{2}+h_{ik}dx^{i}dx^{k}\label{eq:product-metric}\end{equation}
with periodic coordinate $\theta$ on $S^{1}$ and metric $h_{ik}$
on the knot complement $S^{3}\setminus N(K)$. By the ADM formalism
with the lapse $N$ and shift function $N^{i}$ one gets a relation
between the 4-dimensional $R_{K}$ and the 3-dimensional scalar curvature
$R_{(3)}$ (see \cite{MiThWh:73} (21.86) p. 520)\begin{equation}
\sqrt{g_{K}}\, R_{K}\: d^{4}x=N\sqrt{h}\:\left(R_{(3)}+||n||^{2}((tr\mathbf{K})^{2}-tr\mathbf{K}^{2})\right)d\theta\, d^{3}x\label{eq:ADM-splitting}\end{equation}
with the normal vector $n$ and the extrinsic curvature $\mathbf{K}$.
The 4-manifold $S^{1}\times\left(S^{3}\setminus N(K)\right)$ has
a product structure. Because of the diffeomorphism invariance of the
action, we can choose a special coordinate system but get always the
same result. The product metric in $S^{1}\times(S^{3}\setminus N(K))$
allows an embedding $S^{3}\setminus N(K)\hookrightarrow S^{1}\times(S^{3}\setminus N(K))$
in such a manner that the extrinsic curvature do not depend on $\theta$
and it can be chosen to be constant $\mathbf{K}=const.$ One obtains\begin{equation}
S_{exotic}=L_{S^{1}}\cdot\left(\intop_{(S^{3}\setminus N(K))}R_{(3)}\sqrt{h}\, N\, d^{3}x\right)-\lambda\cdot vol(D^{2})\label{eq:relation3-to4dim}\end{equation}
the 3-dimensional\emph{ }Einstein-Hilbert action times the length
$L_{S^{1}}=\intop_{S^{1}}d\theta$ of the circle $S^{1}$. Applying
the identity\[
S^{3}=(S^{3}\setminus N(K))\cup N(K)\]
to the integrals one obtains\[
\intop_{(S^{3}\setminus N(K))}R_{(3)}\sqrt{h}\, N\, d^{3}x+\intop_{N(K)}R_{(3)}\sqrt{h}\, N\, d^{3}x=\intop_{S^{3}}R_{(3)}\sqrt{h}\, N\, d^{3}x=\tilde{\mu}\cdot vol(S^{3})\]
with a constant curvature $\tilde{\mu}>0$ of the 3-sphere and so\begin{equation}
\intop_{(S^{3}\setminus N(K))}R_{(3)}\sqrt{h}\, N\, d^{3}x=\tilde{\mu}\cdot vol(S^{3})-\intop_{N(K)}R_{(3)}\sqrt{h}\, N\, d^{3}x\,.\label{eq:relation-knot-complement-NK}\end{equation}
We remark that the $N(K)=K\times D^{2}$ is defined by an embedding
of the solid torus $S^{1}\times D^{2}$ into the 3-sphere $S^{3}$
and $N(K)$ is the image of this embedding. The boundary of $N(K)$
is a knotted torus $K\times S^{1}$. Especially the total curvature
of the knotted solid torus $N(K)$ and of the knotted torus $S^{1}\times K$
are equal. According to \cite{LangevinRosenberg1976,KuiperMeeks1987}
the integral over $N(K)\subset S^{3}$ is given by the total curvature
of $N(K)$. The greatest lower bound of this curvature is greater
as $8\pi$ for any non-trivial knotted torus (see \cite{KuiperMeeks1987}). 

The integral over $N(K)=K\times D^{2}$ is completely determined by
the boundary (the disk is flatly embedded). It can be calculated as
a term over the boundary $\partial N(K)=K\times S^{1}$, a knotted
torus, i.e. we obtain \begin{eqnarray*}
S_{EH}(N(K)) & = & \intop_{N(K)}R_{(3)}\sqrt{h}N\, d^{3}x=\intop_{\partial(N(K))}X\sqrt{h}d^{2}x\\
 & = & S_{EH}(\partial(N(K)))\end{eqnarray*}
where $X$ is a 2-dimensional expression for the boundary term of
the Einstein-Hilbert action. We will use the same symbol for the 2-dimensional
metric $h$ and its restriction to the boundary submanifold. Now we
are looking for the action at the boundary. As shown by York \cite{York1972},
the fixing of the conformal class of the spatial metric in the ADM
formalism leads to a boundary term which can be also found in the
work of Hawking and Gibbons \cite{GibHaw1977}. Also Ashtekar et.al.
\cite{Ashtekar08,Ashtekar08a} discussed the boundary term in the
Palatini formalism. The main reason for the introduction of the boundary
term came from the Hamiltonian formulation of Einsteins theory. It
has been known since the 1960\textquoteright{}s (see \cite{MiThWh:73}
section 21.4-21.8) that in the Hamiltonian quantization of gravity
it is essential to include boundary terms in the action, as this allows
to define consistently the momentum conjugate to the metric. This
makes it necessary to modify the Einstein-Hilbert action by adding
to it a surface integral term so that the variation of the action
becomes well defined and yields the Einstein field equations. All
these discussions enforce us to choose the following action term at
the boundary $\partial(N(K))$ \[
S_{EH}(\partial(N(K)))=\intop_{\partial(N(K))}H_{\partial}\:\sqrt{h}d^{2}x\]
with $H_{\partial}$ as \emph{mean curvature} of $\partial(N(K))$,
i.e. the trace of the second fundamental form. Of course the mean
curvature do not depend on the coordinate $\theta$ of the circle
$S^{1}$ above. The relation (\ref{eq:relation-knot-complement-NK})
can be extended to the case of $S^{1}\times(S^{3}\setminus N(K))$
by using the product structure above to get\[
\intop_{S^{1}\times(S^{3}\setminus N(K))}R_{(3)}\sqrt{h}\, N\, d\theta d^{3}x=\tilde{\mu}\cdot L_{S^{1}}\cdot vol(S^{3})-\intop_{S^{1}\times N(K)}R_{(3)}\sqrt{h}\, N\, d\theta d^{3}x\,.\]
Therefore we obtain for the exotic action (\ref{eq:EH-action-knot-complement})
\begin{equation}
S_{exotic}=\mu\cdot vol(S^{1}\times S^{3})-\intop_{S^{1}\times\partial(N(K))}H_{\partial}\:\sqrt{h}\, d\theta d^{2}x-\lambda\cdot vol(D^{2})\label{eq:action-splitting-knot-complement}\end{equation}
with the constant $\mu=\tilde{\mu}$ representing the curvature of
$S^{1}\times S^{3}$ which is identical to the curvature of $S^{3}$.
The manifold $S^{1}\times\partial N(K)$ is a knotted 3-torus $T^{3}(K)=K\times S^{1}\times S^{1}$.

Finally we found, that the contribution of the exotic smooth structure
to the usual Einstein-Hilbert action $S_{EM}(M)$ is given by an additive
term -- the exotic action $S_{exotic}$ which is a sum of three Einstein-Hilbert
terms, two over $T^{2}$ and $S^{1}\times S^{3}$ with constant total
curvature and another one over a knotted 3-torus with mean curvature
$H_{\partial}$ \begin{eqnarray}
S_{EH}(M_{K}) & = & S_{EH}(M)+S_{exotic}\label{eq:relation-actions-2a}\\
 & = & S_{EH}(M)+\mu\cdot vol(S^{1}\times S^{3})-\intop_{T^{3}(K)}H_{\partial}\:\sqrt{h}d\theta d^{2}x-\label{eq:relation-actions-2}\\
 &  & -\lambda\cdot vol(D^{2})\nonumber\end{eqnarray}
$T^{3}(K)=S^{1}\times\partial N(K)$ and in usual units ($L_{P}$
Planck length) one has\begin{eqnarray}
\frac{1}{\hbar}S_{EH}(M_{K}) & = & \frac{1}{\hbar}S_{EH}(M)+\frac{1}{L_{P}^{2}}\cdot\mu\cdot vol(S^{1}\times S^{3})-\label{eq:action-functional-for-M}\\
 &  & -\frac{1}{L_{P}^{2}}\left(\intop_{T^{3}(K)}H_{\partial}\:\sqrt{h}d\theta d^{2}x+\lambda\cdot vol(D^{2})\right)\nonumber\,.\end{eqnarray}
In the next section we will show that these exotic action terms have
a common geometrical meaning.

\section{Geometrical Matter\label{sec:Geometrical-interpretation}}

Lets start with the boundary term \[
\intop_{T^{3}(K)}H_{\partial N(K)}\:\sqrt{h}\, d\theta d^{2}x\]
in (\ref{eq:action-functional-for-M}) over the knotted 3-torus $T^{3}(K)=S^{1}\times\partial N(K)$.
The value of this integral depends strongly on the knot $K$ and we
have to say some words about the complexity of knots. %
\begin{figure}
\includegraphics[scale=0.17]{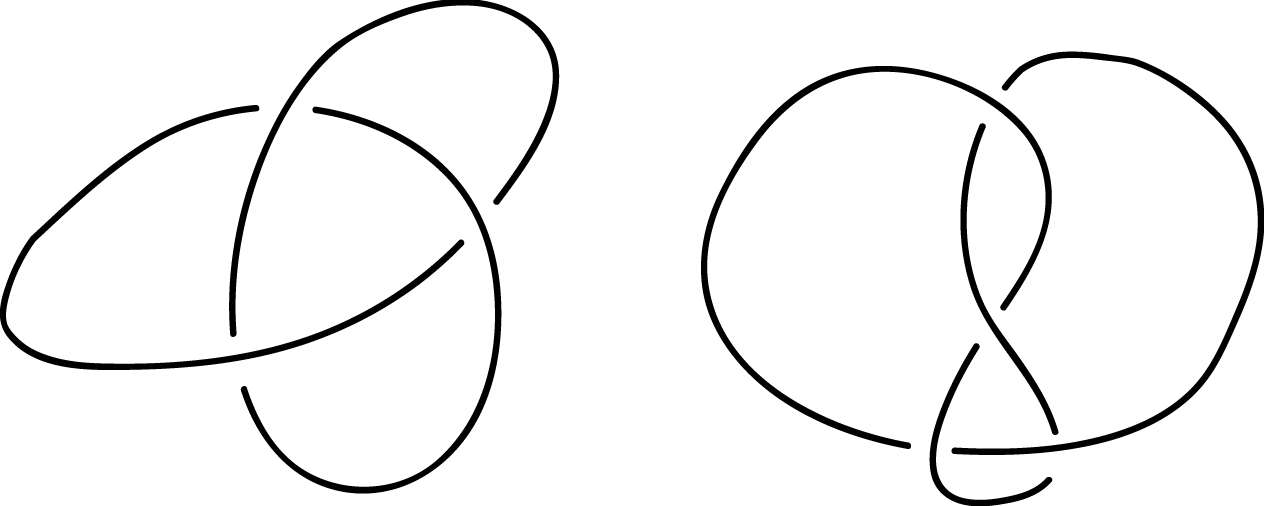} \quad\includegraphics[scale=0.17]{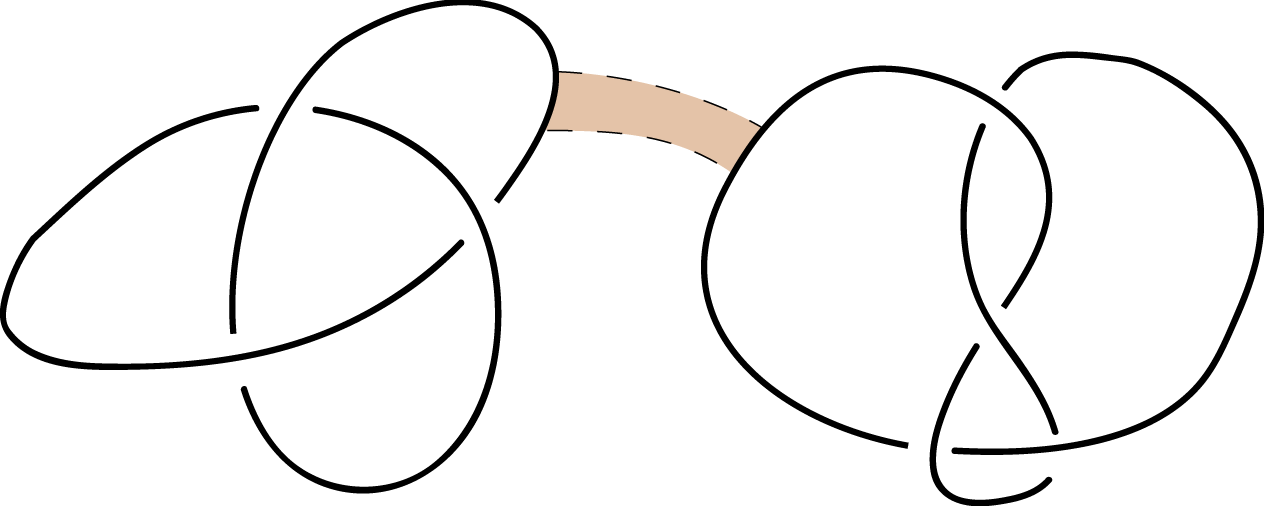}
\quad\includegraphics[scale=0.1]{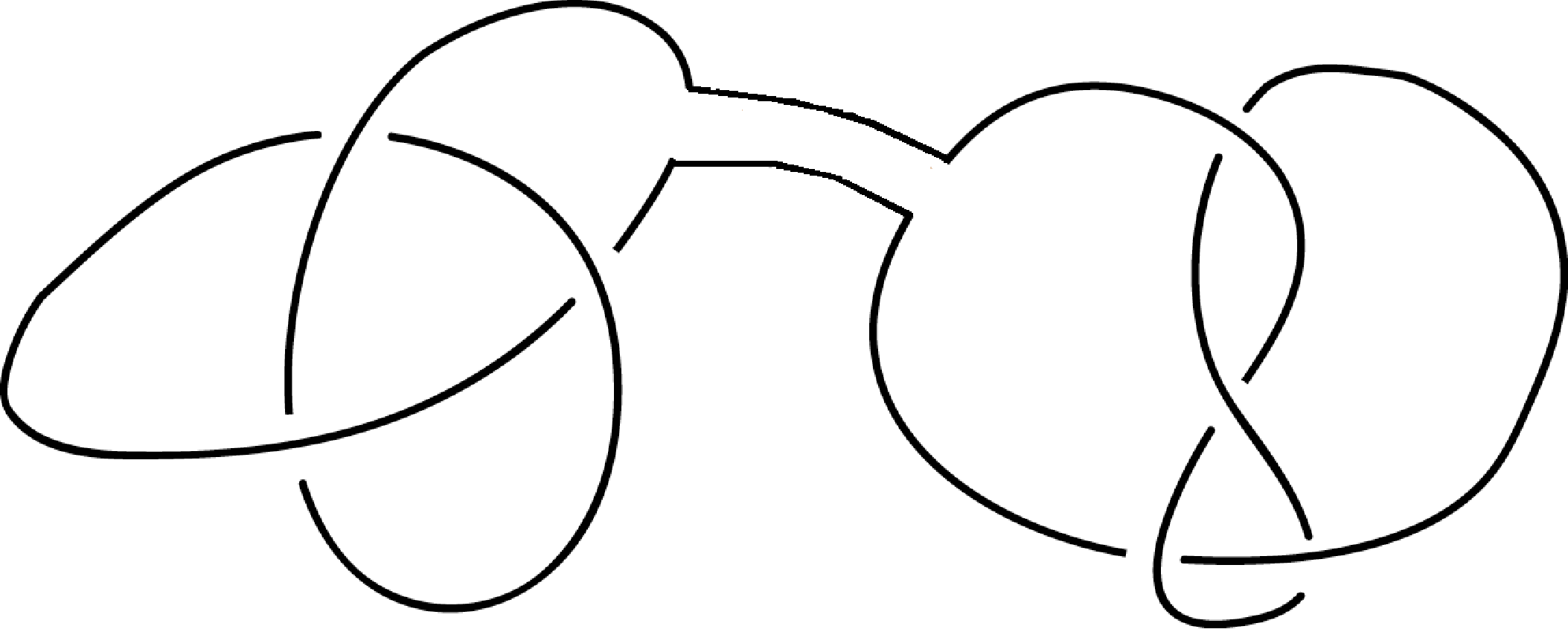}

\caption{Connected sum of knots (left: start, middle:connecting rectangle,
right: join of knots) \label{fig:connected-sum-of-knots}}

\end{figure}
 For that purpose we define the connected sum of two oriented knots:
\begin{itemize}
\item Consider a planar projection of each knot and suppose these projections
are disjoint. (see left figure of Fig. \ref{fig:connected-sum-of-knots})
\item Find a rectangle in the plane where one pair of sides are arcs along
each knot but is otherwise disjoint from the knots and the arcs of
the knots on the sides of the rectangle are oriented around the boundary
of the rectangle in the same direction. (see middle figure of Fig.
\ref{fig:connected-sum-of-knots})
\item Now join the two knots together by deleting these arcs from the knots
and adding the arcs that form the other pair of sides of the rectangle.
(see right figure of Fig. \ref{fig:connected-sum-of-knots})
\end{itemize}
The resulting connected sum knot inherits an orientation consistent
with the orientations of the two original knots. With respect to this
operation there is now a prime factorization: a non-trivial knot is
decomposable if one can represent it by a sum of other non-trivial
knots. A prime knot is a non-trivial knot which cannot be written
as the knot sum of two non-trivial knots. A theorem due to Schubert
\cite{Rol:76} states that every knot can be uniquely expressed as
a connected sum of prime knots. Therefore we have to concentrate on
the prime knots. But one has to respect how this operation affects
the knot complement in the LHS of (\ref{eq:action-splitting-knot-complement}).
For that purpose we consider the integral\[
\intop_{(S^{3}\setminus N(K))}R_{(3)}\sqrt{h}\, N\, d^{3}x\]
over the knot complement $S^{3}\setminus N(K)$. The knot complements
are 3-manifolds with a boundary which is a torus. Like for knots,
there is also a prime decomposition for 3-manifolds. A 3-manifold
$P$ is prime if it cannot be presented as a connected sum in a non-trivial
way, i.e not like $P=P\#S^{3}$. Then \cite{Mil:62}, the prime decomposition
theorem for 3-manifolds states that every compact, orientable 3-manifold
is the connected sum of a unique (up to homeomorphism) collection
of prime 3-manifolds. Knot complements are always prime manifolds.
But it is also possible to split them according to the splitting of
a general knot into a sum of prime knots. Budney \cite{Budney2006}
considers the various operations of knots and its corresponding operation
on the knot complement. Let $C(K)=S^{3}\setminus N(K)$ be the knot
complement for the knot $K$ and assume for $K$a sum \[
K=K_{1}\#K_{2}\]
of prime knots $K_{1},K_{2}$. Then the knot complements admits a
splitting \cite{Budney2006} \[
C(K)=C(K_{1})\cup_{T^{2}}T(K_{1},K_{2})\cup_{T^{2}}C(K_{2})\,.\]
We call $T(K_{1},K_{2})$ the \emph{connecting tube} between the knot
complements $C(K_{1})$ and $C(K_{2})$. This 3-manifold $T(K_{1},K_{2})$
is a so-called graph manifold (or Seifert fibered). In our case, it
can be described by the link complement $T(K_{1},K_{2})=S^{3}\setminus N(H^{2})$
with the so-called key chain link $H^{2}$ (a generalization of a
Hopf link). The connecting tube $T(K_{1},K_{2})$ has a boundary consisting
of three disjoint tori $\partial T(K_{1},K_{2})=T_{1}^{2}\sqcup T_{2}^{2}\sqcup T_{3}^{2}$
(we ignore the orientation) where one of these tori $T_{3}^{2}$ is
the boundary $\partial C(K)=T_{3}^{2}$ of $C(K)$. If we ignore this
boundary (by closing it with a solid torus $T(K_{1},K_{2})\cup_{T_{3}^{2}}(D^{2}\times S^{1})$)
then we have a trivial torus bundle $T^{2}\times[0,1]$ between $T_{1}^{2}$
and $T_{2}^{2}$. The most general operation on knots is the splicing
$J\bowtie K$ (see for the details \cite{Budney2006}) producing examples
of a splitting \[
C(J\bowtie K)=C(K)\cup_{T^{2}}T(K,J)\cup_{T^{2}}C(J)\]
with non-trivial torus bundle $T(K,J)$ (by closing one boundary)
between $C(K)$ and $C(J)$.

The reason why we consider the details of the splitting of knot complements
is that knot complements as prime 3-manifolds are the main objects
of Thurstons geometrization conjecture \cite{Thu:97} (now proven
by Perelman \foreignlanguage{american}{\cite{Per:02,Per:03.1,Per:03.2}})
. This knot complement admits a geometric structure%
\footnote{In 3 dimensions there are 8 geometric structures among them the spherical,
Euclidean and hyperbolic geometry.%
} in the interior $C(K)\setminus\partial C(K)$, i.e. a homogeneous
metric of constant scalar curvature (a Bianchi model in physical notation).
The knots divide into two classes:
\begin{enumerate}
\item hyperbolic knots: the knot complement is a hyperbolic 3-manifold and
\item non-hyperbolic knots: the knot complements admits one of the 7 other
geometric structures.
\end{enumerate}
We would like conjecture that this classification of knots fits well
with the classification of matter into fermions and bosons. Let $K$
be a hyperbolic knot with its hyperbolic complement $C(K)$. Hyperbolic
3-manifolds are subject to a strong restriction called \emph{Mostow
rigidity} \cite{Mos:68}. It states that any diffeomorphism (including
a conformal map) of a hyperbolic 3-manifold is an isometry. Thus geometric
expressions like the volume are topological invariants. This rigidity
is a property which we should expect for fermions. The usual matter
is seen as dust matter (incompressible $p=0$). The scaling behavior
of the energy density $\rho$ for dust matter is determined by the
time-dependent scaling parameter $a$ to be $\rho\sim a^{-3}$. So,
if one represents matter by very small regions in the space equipped
with a geometric structure then this scaling can be generated by an
invariance of these small regions with respect to a rescaling. Mostow
rigidity now singles out the hyperbolic geometry (and the hyperbolic
3-manifold as the corresponding small region) to have the correct
behavior. All other geometries allow a scaling at least along one
direction. The radiation (or interactions represented by bosons as
gauge fields) has a scaling characteristics ($\rho\sim a^{-4}$) like
these geometries. We will discuss the details especially the case
of torus bundles (or connecting tubes) in section \ref{sec:Gauge-group}
more carefully.

For a general knot $K$ (as splicing $K=K_{1}\bowtie K_{2})$ we obtain\begin{eqnarray*}
\intop_{S^{1}\times C(K)}R_{K}\sqrt{g_{K}}\, d^{4}x & = & \intop_{S^{1}\times C(K_{1})}R_{K}\sqrt{g_{K}}\, d^{4}x+\intop_{S^{1}\times T(K_{1},K_{2})}R_{K}\sqrt{g_{K}}\, d^{4}x+\\
 &  & +\intop_{S^{1}\times C(K_{2})}R_{K}\sqrt{g_{K}}\, d^{4}x\,.\end{eqnarray*}
 The action over the knot complements can be written by (\ref{eq:action-splitting-knot-complement})
as integral over the mean curvature. Thus we obtain the contribution
to the action for a general knot:\begin{eqnarray}
\intop_{S^{1}\times C(K)}R_{K}\sqrt{g_{K}}\, d^{4}x & =\mu\cdot vol(S^{1}\times S^{3})- & \sum_{n=1}^{2}\intop_{S^{1}\times\partial N(K_{n})}H_{\partial N(K)}\sqrt{g}d\theta d^{2}x+\nonumber \\
 &  & +\intop_{S^{1}\times T(K_{1},K_{2})}R_{K}\sqrt{g_{K}}\, d^{4}x\label{eq:boundary-term}\end{eqnarray}
And thus we see, that additionally to the usual terms of the Einstein-Hilbert-action
one obtains two types of terms \begin{eqnarray}
 & \intop_{S^{1}\times\partial N(K_{n})}H_{\partial N(K)}\sqrt{g}\, d\theta d^{2}x & ,\label{eq:action fermi}\\
 & \intop_{S^{1}\times T(K_{1},K_{2})}R_{K}\sqrt{g_{K}}\, d^{4}x & ,\label{eq:action bose}\end{eqnarray}
in (\ref{eq:action-functional-for-M}). Both types of integrals describe
the immersion of certain submanifolds into the 3-space. In the following
we will show that these terms can be interpreted as the action of
(fermionic) spinor fields as well as of (bosonic) gauge fields.

\section{Dirac action\label{sec:Dirac-action}}

The action (\ref{eq:action fermi}) above is completely determined
by the knotted torus $\partial N(K)=K\times S^{1}$ and its mean curvature
$H_{\partial N(K)}$. This knotted torus is an immersion of a torus
$S^{1}\times S^{1}$ into $\mathbb{R}^{3}$. The well-known \emph{Weierstrass
representation} can be used to describe this immersion. As proved
in \cite{SpinorRep1996,Friedrich1998} there is an equivalent representation
via spinors. This so-called \emph{Spin representation} of a surface
gives back an expression for $H_{\partial N(K)}$ and the Dirac equation
as geometric condition on the immersion of the surface. As we will
show below, the term (\ref{eq:action fermi}) can be interpreted as
Dirac action of a spinor field.

\subsection{Weierstrass and spin representation of immersed submanifolds\label{subsubsec:Spin-representation-disk}}

In this subsection we describe the theory of immersions using spinors.
The theory will be presented stepwise. We start with a toy model of
an immersion of a surface into the 3-dimensional Euclidean space.
Then we discuss how this map can be extended to an immersion of a
3-manifold into a 4-manifold. 

Let $f:M^{2}\to\mathbb{R}^{3}$ be a smooth map of a Riemannian surface
with injective differential $df:TM^{2}\to T\mathbb{R}^{3}$, i.e.
an immersion. In the \emph{Weierstrass representation} one expresses
a \emph{conformal minimal} immersion $f$ in terms of a holomorphic
function $g\in\Lambda^{0}$ and a holomorphic 1-form $\mu\in\Lambda^{1,0}$
as the integral\[
f=Re\left(\int(1-g^{2},i(1+g^{2}),2g)\mu\right)\ .\]
An immersion of $M^{2}$ is conformal if the induced metric $g$ on
$M^{2}$ has components\[
g_{zz}=0=g_{\bar{z}\bar{z}}\,,\: g_{z\bar{z}}\not=0\]
and it is minimal if the surface has minimal volume. Now we consider
a spinor bundle $S$ on $M^{2}$ (i.e. $TM^{2}=S\otimes S$ as complex
line bundles) and with the splitting\[
S=S^{+}\oplus S^{-}=\Lambda^{0}\oplus\Lambda^{1,0}\]
Therefore the pair $(g,\mu)$ can be considered as spinor field $\varphi$
on $M^{2}$. Then the Cauchy-Riemann equation for $g$ and $\mu$
is equivalent to the Dirac equation $D\varphi=0$. The generalization
from a conformal minimal immersion to a conformal immersion was done
by many authors (see the references in \cite{Friedrich1998}) to show
that the spinor $\varphi$ now fulfills the Dirac equation\begin{equation}
D\varphi=H\varphi\label{eq:conformal-immersion-Dirac}\end{equation}
where $H$ is the mean curvature (i.e. the trace of the second fundamental
form). The minimal case is equivalent to the vanishing mean curvature
$H=0$ recovering the equation above. Friedrich \cite{Friedrich1998}
uncovered the relation between a spinor $\Phi$ on $\mathbb{R}^{3}$
and the spinor $\varphi=\Phi|_{M^{2}}$: if the spinor $\Phi$ fulfills
the Dirac equation $D\Phi=0$ then the restriction $\varphi=\Phi|_{M^{2}}$
fulfills equation (\ref{eq:conformal-immersion-Dirac}) and $|\varphi|^{2}=const$.
Therefore we obtain\begin{equation}
H=\bar{\varphi}D\varphi\label{eq:mean-curvature-surface}\end{equation}
with $|\varphi|^{2}=1$. 

After this exercise we are ready to consider the integral (\ref{eq:action fermi}).
Here we have an immersion of a torus $I:T^{2}=S^{1}\times S^{1}\to\mathbb{R}^{3}$
with image the knotted torus $im(I)=T(K)=K\times S^{1}$ that is the
boundary $\partial N(K)$ of $N(K)$. This immersion $I$ can be defined
by a spinor $\varphi$ on $T^{2}$ fulfilling the Dirac equation\begin{equation}
D\varphi=H\varphi\label{eq:2D-Dirac}\end{equation}
with $|\varphi|^{2}=1$ (or an arbitrary constant) (see Theorem 1
of \cite{Friedrich1998}). As discussed above a spinor bundle over
a surface splits into two sub-bundles $S=S^{+}\oplus S^{-}$ with
the corresponding splitting of the spinor $\varphi$ in components\[
\varphi=\left(\begin{array}{c}
\varphi^{+}\\
\varphi^{-}\end{array}\right)\]
and we have the Dirac equation\[
D\varphi=\left(\begin{array}{cc}
0 & \partial_{z}\\
\partial_{\bar{z}} & 0\end{array}\right)\left(\begin{array}{c}
\varphi^{+}\\
\varphi^{-}\end{array}\right)=H\left(\begin{array}{c}
\varphi^{+}\\
\varphi^{-}\end{array}\right)\]
with respect to the coordinates $(z,\bar{z})$ on $T^{2}$. 

In dimension 3, the spinor bundle has the same fiber dimension as
the spinor bundle $S$ (but without a splitting $S=S^{+}\oplus S^{-}$into
two sub-bundles). Now we define the extended spinor $\phi$ over the
3-torus $T^{3}=S^{1}\times S^{1}\times S^{1}$ via the restriction
$\phi|_{T^{2}}=\varphi$. The spinor $\phi$ is constant along the
normal vector $\partial_{N}\phi=0$ fulfilling the 3-dimensional Dirac
equation\begin{equation}
D^{3D}\phi=\left(\begin{array}{cc}
\partial_{N} & \partial_{z}\\
\partial_{\bar{z}} & -\partial_{N}\end{array}\right)\phi=H\phi\label{eq:Dirac-equation-3D}\end{equation}
induced from the Dirac equation (\ref{eq:2D-Dirac}) via restriction
and where $|\phi|^{2}=const.$ Especially one obtains for the mean
curvature of the knotted 3-torus $T^{3}(K)=K\times S^{1}\times S^{1}$
(up to a constant from $|\phi|^{2}$)\begin{equation}
H=\bar{\phi}D^{3D}\phi\,.\label{eq:mean-curvature-3D}\end{equation}

\subsection{The Dirac action in 3 dimensions and the 4-dimensional Dirac equation}

By using the relation (\ref{eq:mean-curvature-3D}) above we obtain
for the integral (\ref{eq:action fermi})\begin{equation}
\intop_{S^{1}\times\partial N(K_{n})}H_{\partial N(K)}\sqrt{g}d\theta d^{2}x=\intop_{S^{1}\times\partial N(K)}\bar{\phi}D^{3D}\phi\:\sqrt{g}\, d\theta d^{2}x\label{eq:3D-action-fermion}\end{equation}
i.e. the Dirac action on the knotted 3-torus $S^{1}\times\partial N(K)=K\times S^{1}\times S^{1}=T^{3}(K)$.
But that is not the expected result, we obtain only a 3-dimensional
Dirac action leaving us with the question to extend the action to
four dimensions. 

Let $\iota:T^{3}\hookrightarrow M$ be an immersion of the 3-torus
$\Sigma=T^{3}$ into the 4-manifold $M$ with the normal vector $\vec{N}$.
At this stage one can consider an arbitrary 3-manifold $\Sigma$ instead
of the 3-torus. The spin bundle $S_{M}$ of the 4-manifold splits
into two sub-bundles $S_{M}^{\pm}$ where one subbundle, say $S_{M}^{+},$
can be related to the spin bundle $S_{\Sigma}$ of the 3-manifold.
Then the spin bundles are related by $S_{\Sigma}=\iota^{*}S_{M}^{+}$
with the same relation $\phi=\iota_{*}\Phi$ for the spinors ($\phi\in\Gamma(S_{\Sigma})$
and $\Phi\in\Gamma(S_{M}^{+})$). Let $\nabla_{X}^{M},\nabla_{X}^{\Sigma}$
be the covariant derivatives in the spin bundles along a vector field
$X$ as section of the bundle $T\Sigma$. Then we have the formula\begin{equation}
\nabla_{X}^{M}(\Phi)=\nabla_{X}^{\Sigma}\phi-\frac{1}{2}(\nabla_{X}\vec{N})\cdot\vec{N}\cdot\phi\label{eq:covariant-derivative-immersion}\end{equation}
with the obvious embedding $\phi\mapsto\left(\begin{array}{c}
\phi\\
0\end{array}\right)=\Phi$ of the spinor spaces. The expression $\nabla_{X}\vec{N}$ is the
second fundamental form of the immersion where the trace $tr(\nabla_{X}\vec{N})=2H$
is related to the mean curvature $H$. Then from (\ref{eq:covariant-derivative-immersion})
one obtains a similar relation between the corresponding Dirac operators\begin{equation}
D^{M}\Phi=D^{3D}\phi-H\phi\label{eq:relation-Dirac-3D-4D}\end{equation}
with the Dirac operator $D^{3D}$ defined via (\ref{eq:Dirac-equation-3D}).
Together with equation (\ref{eq:Dirac-equation-3D}) we obtain\begin{equation}
D^{M}\Phi=0\label{eq:Dirac-equation-4D}\end{equation}
i.e. $\Phi$ is a parallel spinor.

\subsection{The extension to the 4-dimensional Dirac action}

Above we obtained a relation (\ref{eq:relation-Dirac-3D-4D}) between
a 3-dimensional spinor $\phi$ on the 3-manifold $\Sigma=T^{3}$ fulfilling
a Dirac equation $D^{\Sigma}\phi=H\phi$ (determined by the immersion
$\Sigma\to M$ into a 4-manifold $M$) and a 4-dimensional spinor
$\Phi$ on a 4-manifold $M$ with fixed chirality ($\in\Gamma(S_{M}^{+})$
or $\in\Gamma(S_{M}^{-})$) fulfilling the Dirac equation $D^{M}\Phi=0$.
At first we consider the variation\begin{equation}
\delta\intop_{S^{1}\times\partial N(K)}\bar{\phi}D^{3D}\phi\:\sqrt{g}\, d\theta d^{2}x=0\label{eq:3D-variation}\end{equation}
of the 3-dimensional action leading to the Dirac equations\begin{equation}
D^{3D}\phi=0\quad D^{3D}\bar{\phi}=0\label{eq:3D-Dirac-equation}\end{equation}
or to \[
H=0\,,\]
a characterization of the immersion $S^{1}\times\partial N(K)$ of
the 3-torus $T^{3}$ with minimal mean curvature. This variation can
be understood as a variation of the (conformal) immersion. In contrast,
the extension of the spinor $\phi$ (as solution of (\ref{eq:3D-Dirac-equation}))
to the 4-dimensional spinor $\Phi$ by using the embedding\begin{equation}
\Phi=\left(\begin{array}{c}
\phi\\
0\end{array}\right)\label{eq:embedding-spinor-3D-4D}\end{equation}
can be only seen as immersion, if (and only if) the 4-dimensional
Dirac equation\[
D^{M}\Phi=0\]
on $M$ is fulfilled (using relation (\ref{eq:relation-Dirac-3D-4D})).
This Dirac equation is obtained by varying the action\begin{equation}
\delta\intop_{M}\bar{\Phi}D^{M}\Phi\sqrt{g}\: d^{4}x=0\label{eq:4D-variation}\end{equation}
Importantly, this variation has a different interpretation in contrast
to varying the 3-dimensional action. Both variations look very similar.
But in (\ref{eq:4D-variation}) we vary over smooth maps $\Sigma=T^{3}\to M$
which are not conformal immersions (i.e. represented by spinors $\Phi$
with $D^{M}\Phi\not=0$). Only the choice of the extremal action selects
the conformal immersion among other smooth maps. Especially the spinor
$\Phi$ (as solution of the 4-dimensional Dirac equation) is localized
at the immersed 3-manifold $\Sigma$ (with respect to the embedding
(\ref{eq:embedding-spinor-3D-4D})). The 3-manifold $\Sigma$ moves
along the normal vector (see the relation (\ref{eq:covariant-derivative-immersion})
between the covariant derivatives representing a parallel transport).

Therefore the 3-dimensional action (\ref{eq:3D-action-fermion}) can
be extended to the whole 4-manifold (but for a spinor $\Phi$ of fixed
chirality). Especially we have a unique fermionic action on the manifold
$M$. By combining the action (\ref{eq:action-functional-for-M})
with (\ref{eq:boundary-term}) and ignoring the action (\ref{eq:action bose})
of the connecting tubes, one obtains the pure fermionic action on
$M$ \begin{equation}
\intop_{M}(R+\bar{\Phi}D^{M}\Phi)\sqrt{g}d^{4}x\label{eq:fermionic-action}\end{equation}
The action (\ref{eq:fermionic-action}) is the usual Einstein-Hilbert
action for a Dirac field $\Phi$ as source. What about the mass term?
In our scheme there is one possible way to do it: using the constant
length $|\Phi|^{2}=const.$ of the spinor, we can introduce the scalar
curvature $R_{\Gamma}$ of an additional 3-manifold $\Gamma$ with
constant curvature coupled to the spinor. Then we obtain \begin{equation}
\intop_{M}\bar{\Phi}(D^{M}-m)\Phi\sqrt{g}\: d^{4}x\label{eq:mass-term-fermion}\end{equation}
with $m=-R_{\Gamma}$ and $\Gamma\subset M$. But we already have
a natural choice for this manifold, the 3-sphere $\Gamma=S^{3}$ as
the embedding space for the knotted torus $\partial N(K)=K\times S^{1}$.
Then the knotted 3-torus $T^{3}(K)=K\times S^{1}\times S^{1}$ is
given by an embedding of the 3-torus $T^{3}$ into $S^{1}\times S^{3}$.
Therefore as a conjecture the term $\mu\cdot vol(S^{1}\times S^{3})$
in the action (\ref{eq:action-functional-for-M}) can be interpreted
as mass term for the fermions. Especially we obtain\[
\intop_{M}m\bar{\Phi}\Phi\sqrt{g}\: d^{4}x=\mu\cdot vol(S^{1}\times S^{3})\]
having the correct sign in the action (\ref{eq:mass-term-fermion}).
But at the moment we have no idea how to generate realistic masses
from this idea.

\section{Gauge field action\label{sec:Gauge-field-action}}

Now we will discuss the second term (\ref{eq:action bose})\[
\intop_{S^{1}\times T(K_{1},K_{2})}R_{K}\sqrt{g_{K}}\, d^{4}x\]
Using the product metric (\ref{eq:product-metric}) and the splitting
(\ref{eq:ADM-splitting}) as well the relation $\mathbf{K}=const.$
to rewrite this integral \begin{eqnarray*}
S_{EH}(S^{1}\times T(K_{1},K_{2})) & = & \intop_{S^{1}\times T(K_{1},K_{2})}R_{K}\sqrt{g_{K}}\, d^{4}x\\
 & = & L_{S^{1}}\intop_{T(K_{1},K_{2})}R_{(3)}\sqrt{h}\, N\, d^{3}x\end{eqnarray*}
As shown by Witten \cite{Wit:89.2,Wit:89.3,Wit:91.2}, the action\[
\intop_{T(K_{1},K_{2})}R_{(3)}\sqrt{h}\, N\, d^{3}x=L\cdot CS(T(K_{1},K_{2}),A)\]
is related to the Chern-Simons action $CS(T(K_{1},K_{2}),A)$ (defined
in the appendix \ref{sec:Chern-Simons-invariant}) and we obtain for
the action $S_{EH}(S^{1}\times T(K_{1},K_{2}))$\begin{equation}
\intop_{S^{1}\times T(K_{1},K_{2})}R_{K}\sqrt{g_{K}}\, d^{4}x=L_{S^{1}}\cdot L\cdot CS(T(K_{1},K_{2}),A)\label{eq:CS-relation}\end{equation}
with respect to the (Levi-Civita) connection $A$ and the length $L$.
For the 3-manifold $T(K_{1},K_{2})$, there is a 4-manifold $M_{T}$
with $\partial M_{T}=T(K_{1},K_{2})$ (take for instance $M_{T}=T(K_{1},K_{2})\times[0,1)\subset T(K_{1},K_{2})\times S^{1}$).
By using the Stokes theorem (see (\ref{eq:stokes-CS}) in the appendix
\ref{sec:Chern-Simons-invariant}) we obtain\[
S_{EH}(M_{T})=\intop_{M_{T}}tr(F\wedge F)\]
with the curvature $F=DA$, i.e. the action is the (topological) Pontrjagin
class of the 4-manifold $M_{T}$. But $T(K_{1},K_{2})$ is a manifold
with boundary and thus the variation of the action do not vanish.
From the formal point of view, the curvature 2-form $F=DA$ is generated
by a $SO(3,1)$ connection $A$ in the frame bundle, which can be
lifted uniquely to a $SL(2,\mathbb{C})$- (Spin-) connection. According
to the Ambrose-Singer theorem, the components of the curvature tensor
are determined by the values of holonomy which is in general a subgroup
of $SL(2,\mathbb{C})$.

Thus we start with a suitable curvature 2-form $F=DA$ with values
in the Lie algebra $\mathfrak{g}$ of the Lie group $G$ as subgroup
of the $SL(2,\mathbb{C})$. The variation of the Chern-Simons action
(\ref{eq:CS-relation}) gets flat connections $DA=0$ as solutions.
The flow of solutions $A(t)$ in $T(K_{1},K_{2})\times[0,1]$ (parametrized
by the variable $t$, the ''time'') between the flat connection $A(0)$
in $T(K_{1},K_{2})\times\left\{ 0\right\} $ to the flat connection
$A(1)$ in $T(K_{1},K_{2})\times\left\{ 1\right\} $ will be given
by the gradient flow equation (see \cite{Flo:88} for instance)\begin{equation}
\frac{d}{dt}A(t)=\pm*F(A)=\pm*DA\label{eq:gradient-flow}\end{equation}
where the coordinate $t$ is normal to $T(K_{1},K_{2})$. Therefore
we are able to introduce a connection $\tilde{A}$ in $T(K_{1},K_{2})\times[0,1]$
so that the covariant derivative in $t$-direction agrees with $\partial/\partial t$.
Then we have for the curvature $\tilde{F}=D\tilde{A}$ where the fourth
component is given by $\tilde{F}_{4\mu}=d\tilde{A}_{\mu}/dt$. Thus
we will get the instanton equation with (anti-)self-dual curvature\[
\tilde{F}=\pm*\tilde{F}\,.\]
It follows\[
S_{EH}([0,1]\times T(K_{1},K_{2}))=\intop_{T(K_{1},K_{2})\times[0,1]}tr(\tilde{F}\wedge\tilde{F})=\pm\intop_{T(K_{1},K_{2})\times[0,1]}tr(\ \tilde{F}\wedge*\tilde{F})\,,\]
i.e. the action of the gauge field. The whole procedure remains true
for an extension of the ''time'', i.e.\begin{equation}
S_{EH}(\mathbb{R}\times T(K_{1},K_{2}))=\pm\intop_{T(K_{1},K_{2})\times\mathbb{R}}tr(\ \tilde{F}\wedge*\tilde{F})\,.\label{eq:gauge-action-tubes}\end{equation}

\subsection{Extension to the 4-dimensional action}

The gauge field action (\ref{eq:gauge-action-tubes}) is only defined
along the tubes $T(K_{1},K_{2})$. For the extension of the action
to the whole 4-manifold $M$, we need some non-trivial facts from
the theory of 3-manifolds. At first the tubes $T(K_{1},K_{2})$ are
so-called graph-manifolds. The decomposition of a general irreducible
3-manifold is given by the Thick-Thin decomposition: the thick part
is a collection of hyperbolic 3-manifolds whereas the thin part is
the set of graph manifolds. This decomposition is a result of the
Mostow rigidity theorem \cite{Mos:68}, i.e. every conformal transformation
(especially a scaling) of a hyperbolic manifold is an isometry. Therefore
there exists a scaling which makes the thin part smaller but does
not change the thick part. Of course the tubes $T(K_{1},K_{2})$ are
contained in the thin part. We contract $T(K_{1},K_{2})$ to thin
tubes connecting the thick parts. Conversely one also finds a scaling
so that the thin part becomes large (but the thick part has the same
size). Thus we can interpret the curvature $\tilde{F}$ of the thin
part as field located between the thick part. The thick part can be
interpreted as fermions (see above). Then the action integral of the
bosons can be written as\[
\intop_{M\setminus vol(fermion)}tr(\tilde{F}\wedge*\tilde{F})\]
i.e. like the integral (\ref{eq:gauge-action-tubes}) considered over
$M\setminus vol(fermions)$, the spacetime except the thick part (extended
along the time axis). In the point-particle case we can extend this
field to the whole manifold $M$. Then we obtain the gauge field action\begin{equation}
\intop_{M}tr(\tilde{F}\wedge*\tilde{F})\label{eq:gauge-field-action}\end{equation}
In the action (\ref{eq:action-functional-for-M}) we have two constant
terms $\mu\cdot vol(S^{1}\times S^{3})$ and $\lambda\cdot vol(D^{2})$
(coming from the integral (\ref{eq:flat-torus})). The first constant
was interpreted as a mass term of the fermion. But for the second
constant $\lambda\cdot vol(D^{2})$ we will conjecture a relation
to the \emph{cosmological constant} $\Lambda$ usually defined by
\[
\intop_{M}\Lambda\,\sqrt{g}\, d^{4}x=\Lambda\cdot vol(M)\:.\]
But the comparison with the term $\lambda$ in the action (\ref{eq:action-functional-for-M})
gives\begin{equation}
\Lambda=\frac{\lambda\cdot vol(D^{2})}{vol(M)}\:.\label{eq:cosmo-constant}\end{equation}
Finally we summarize all terms for the whole action of the proposed
geometrical model of matter (\ref{eq:action-functional-for-M}) \begin{equation}
S(M)=\intop_{M}\left(R-\Lambda+\sum_{n}(\bar{\Phi}(D^{M}-m)\Phi)_{n}\right)\sqrt{g}\: d^{4}x+\intop_{M}tr(\tilde{F}\wedge*\tilde{F})\,.\label{eq:action-on-M-standard-model}\end{equation}
showing a combined Dirac-gauge-field coupled to the Einstein-Hilbert
action.

\section{Gauge group\label{sec:Gauge-group}}

In the last section we have seen, that the connecting tubes of the
geometrical model can be interpreted as a gauge field. We will now
discuss the possible gauge group of the obtained field. The gauge
field in the action (\ref{eq:gauge-field-action}) has values in the
Lie algebra of the maximal compact subgroup $SU(2)$ of $SL(2,\mathbb{C})$.
But in the derivation of the action, we used the connecting tube $T(K_{1},K_{2})$
between two tori which is a cobordism. This cobordism $T(K_{1},K_{2})$
is also known as torus bundle (see \cite{Calegari2007} Theorem 1.15)
which can be always decomposed into three elementary pieces -- \emph{finite
order}, \emph{Dehn twist} and \emph{Anosov map}%
\footnote{The details of the construction is not important for the following
discussion.%
}. The idea of this construction is very simple: one starts with two
trivial cobordisms $T^{2}\times[0,1]$ and glue them together by using
a diffeomorphism $g:T^{2}\to T^{2}$ which we call \emph{gluing diffeomorphism}.
From the geometrical point of view, we have to distinguish between
three different types of torus bundles. The three types of torus bundles
are distinguished by the splitting of the tangent bundle: 
\begin{itemize}
\item finite order (orders $2,3,4,6$): the tangent bundle is 3-dimensional 
\item Dehn-twist (left/right twist): the tangent bundle is a sum of a 2-dimensional
and a 1-dimensional bundle
\item Anosov: the tangent bundle is a sum of three 1-dimensional bundles.
\end{itemize}
Following Thurston's geometrization program (see \cite{Thu:97}),
these three torus bundles are admitting a geometric structure, i.e.
it has a metric of constant curvature. Apart from this geometric properties,
all torus bundles are determined by the gluing diffeomorphism $g:T^{2}\to T^{2}$
determing also the fundamental group of the torus bundle. Therefore
this gluing diffeomorphism has also influence on the structure of
the diffeomorphism group of the torus bundle which will be discussed
now. From the physical point of view, we have two types of diffeomorphisms:
local and global. Any coordinate transformation can be described by
an infinitesimal or local diffeomorphism (coordinate transformation).
In contrast there are global diffeomorphisms like an orientation reversing
diffeomorphism. Two diffeomorphisms not connected via a sequence of
local diffeomorphism are part of different connecting components of
the diffeomorphism group, i.e. the set of isotopy classes $\pi_{0}(Diff(M))$
(also called the mapping class group). Isotopy classes are important
to understand the configuration space topology of general relativity
(see Giulini \cite{Giulini94}). Consider two different isotopy classes
of a given 3-manifold. By definition, the two 3-manifolds cannot be
connected by a sequence of local diffeomorphisms (local coordinate
transformtions). Therefore two different isotopy classes represent
two different physical situations, see \cite{Giulini09} for the relation
of isotopy classes to particle properties like spin. In case of the
torus bundle we consider the isotopy classes $\pi_{0}(Diff(M,\partial M))$
relativ to the boundary represented by the automorphisms of the fundamental
group. Using the geometrization program, we obtain a relation between
the isotopy classes $\pi_{0}(Diff(M,\partial M))$ and the isometry
classes (connecting components of the isometry group) with respect
to the geometric structure of the torus bundle (see for instance \cite{KalliongisMcCullough1996,HatcherMcCullough1997}).
Then the isotopy classes of the torus bundles are given by 
\begin{itemize}
\item finite order: 2 isotopy classes (= no/even twist or odd twist)
\item Dehn-twist: 2 isotopy classes (= left or right Dehn twists)
\item Anosov: 8 isotopy classes (= all possible orientations of the three
line bundles forming the tangent bundle)
\end{itemize}
From the geometrical point of view, we can rearrange the scheme above:
\begin{itemize}
\item torus bundle with no/even twists: 1 isotopy class
\item torus bundle with twist (Dehn twist or odd finite twist): 3 isotopy
classes
\item torus bundle with Anosov map: 8 isotopy classes
\end{itemize}
This information puts a starting point of the discussion how to derive
the gauge group. Given a Lie group $G$ with Lie algebra $\mathfrak{g}$.
The rank of $\mathfrak{g}$ is the dimension of the maximal abelian
subalgebra, also called Cartan algebra. It is the same as the dimension
of the maximal torus $T^{n}\subset G$. The curvature $F$ of the
gauge field takes values in the adjoint representation of the Lie
algebra and the action $tr(F\wedge*F)$ forms an element of the Cartan
subalgebra (the Casimir operator). But each isotopy class contributes
to action and therefore we have to take the sum over the isotopy classes.
Let $t_{a}$ the generator in the adjoint representation, then we
obtain for the Lie algebra part of the action $tr(F\wedge*F)$
\begin{itemize}
\item torus bundle with no twists: 1 isotopy class with $t^{2}$
\item torus bundle with twist: 3 isotopy classes with $t_{1}^{2}+t_{2}^{2}+t_{3}^{2}$
\item torus bundle with Anosov map: 8 isotopy classes with $\sum_{a=1}^{8}t_{a}^{2}$.
\end{itemize}
The Lie algebra with one generator $t$ corresponds uniquely to the
Lie group $U(1)$ where the 3 generators $t_{1},t_{2},t_{3}$ form
the Lie algebra of the $SU(2)$ group. Then the last case with 8 generators
$t_{a}$ have to correspond to the Lie algebra of the $SU(3)$ group.
We remark the similarity with an idea from brane theory: $n$ parallel
branes (each decorated with an $U(1)$ theory) are described by an
$U(n)$ gauge theory (see \cite{GiveonKutasov99}). Finally we obtain
the maximal group $U(1)\times SU(2)\times SU(3)$ as gauge group for
all possible torus bundle (in the model: connecting tubes between
the solid tori).

At the end we will speculate about the identification of the isotopy
classes for the torus bundle with the vector bosons in the gauge field
theory. Obviously the isotopy class of the torus bundle with no twist
must be the photon. Then the isotopy class of the other bundle of
finite order should be identified with the $Z^{0}$ boson and the
two isotopy classes of the Dehn twist bundles are the $W^{\pm}$ bosons.
We remark that this scheme contains automatically the mixing between
the photon and the $Z^{0}$ boson (the corresponding torus bundle
are both of finite order). The isotopy classes of the Anosov map bundle
have to correspond to the 8 gluons. We know that this approach left
open many questions (like symmetry breaking, Higgs boson etc.).

\section{Discussion}

At the end of the paper we will summarize our assumptions and results.
We started with a smooth 4-manifold $M$ admitting an exotic smoothness
structure $M_{K}$. This smoothness structure is constructed by using
knot surgery. Then we discuss the general properties like the existence
of a Lorentz metric and global hyperbolicity. Beginning with section
\ref{sec:Matter-as-exotic-space} we considered the Einstein-Hilbert
action on $M_{K}$ and the decomposition \[
M=\left(M\setminus N(T^{2})\right)\cup_{T^{2}}\left(S^{1}\times\left(S^{3}\setminus N(K)\right)\right)\]
Because of the diffeomorphism invariance of the action, one can split
the Einstein-Hilbert action like (\ref{eq:relation-action-1}) leading
to the relation (\ref{eq:relation-action-knot-surgery}). Then we
were able to define an action over the knot complement to identify
two contributions: knotted tori and connecting tubes between two tori. 
\begin{enumerate}
\item knotted solid torus: As shown in section \ref{sec:Dirac-action},
a knotted solid torus can be described by a spinor so that the mean
curvature is the Dirac action of this spinor. This action over the
3-dimensional boundary can be extended to the whole 4-manifold (\ref{eq:fermionic-action}).
\item connecting tube: We discussed this case in section \ref{sec:Gauge-field-action}
using special properties of the tube as cobordism between two tori.
Finally we obtained the Yang-Mills action (\ref{eq:gauge-field-action}).
\end{enumerate}
We finish the paper with a conjecture about the gauge group. The connecting
tubes can be identified with torus bundles which are classified. The
three possible types of torus bundles were identified with three interactions
to get the gauge group $U(1)\times SU(2)\times SU(3)$. Finally we
can support the conjecture that exotic smoothness generates fermionic
and bosonic fields\emph{.}
\begin{acknowledgements}
We thank A. Ashtekar for his remarks about the boundary term of the
Einstein-Hilbert action. Many thanks to Carl H. Brans for many discussions
about the physics of exotic 4-manifolds. The section about global
hyperbolicity is mainly inspired by the discussion with Miguel S\'anchez.
Many thanks for his helpful remarks. We acknowledge the critical but
helpful remarks from the referees. 
\end{acknowledgements}
\appendix
{

\section{Knot complement\label{sec:Knot-complement} }

Let $K:S^{1}\to S^{3}$ be an embedding of the circle into the 3-sphere,
i.e. a knot $K$. We define by $N(K)=D^{2}\times K$ a thickened knot
or a knotted solid torus. The knot complement $S^{3}\setminus N(K)$
results in cutting $N(K)$ off from the 3-sphere $S^{3}$. Then one
obtains a 3-manifold with boundary $\partial(S^{3}\setminus N(K))=T^{2}$.
The properties of the knot complement depend strongly on the properties
of the knot $K$. So, the fundamental group $\pi_{1}(S^{3}\setminus N(K))$
is also denoted as knot group. In contrast, the homology group $H_{1}(S^{3}\setminus N(K))=\mathbb{Z}$
don't depend on the knot.

\section{Chern-Simons invariant\label{sec:Chern-Simons-invariant}}

Let $P$ be a principal $G$ bundle over the 4-manifold $M$ with
$\partial M\not=0$. Furthermore let $A$ be a connection in $P$
with the curvature \[
F_{A}=dA+A\wedge A\]
and Chern class\[
C_{2}=\frac{1}{8\pi^{2}}\int\limits _{M}tr(F_{A}\wedge F_{A})\]
for the classification of the bundle $P$. By using the Stokes theorem
we obtain \begin{equation}
\int\limits _{M}tr(F_{A}\wedge F_{A})=\int\limits _{\partial M}tr(A\wedge dA+\frac{2}{3}A\wedge A\wedge A)\label{eq:stokes-CS}\end{equation}
with the Chern-Simons invariant \begin{equation}
CS(\partial M,A)=\frac{1}{8\pi^{2}}\int\limits _{\partial M}tr(A\wedge dA+\frac{2}{3}A\wedge A\wedge A)\:.\label{CS-invariante}\end{equation}
Now we consider the gauge transformation $A\rightarrow g^{-1}Ag+g^{-1}dg$
and obtain\[
CS(\partial M,g^{-1}Ag+g^{-1}dg)=CS(\partial M,A)+k\]
with the winding number \[
k=\frac{1}{24\pi^{2}}\int\limits _{\partial M}(g^{-1}dg)^{3}\in\mathbb{Z}\]
of the map $g:M\rightarrow G$. Thus the expression \[
CS(\partial M,A)\bmod1\]
is an invariant, the Chern-Simons invariant. Now we will calculate
this invariant. For that purpose we consider the functional (\ref{CS-invariante})
and its first variation vanishes \[
\delta CS(\partial M,A)=0\]
because of the topological invariance. Then one obtains the equation
\[
dA+A\wedge A=0\:,\]
i.e. the extrema of the functional are the connections of vanishing
curvature. The set of these connections up to gauge transformations
is equal to the set of homomorphisms $\pi_{1}(\partial M)\rightarrow SU(2)$
up to conjugation. Thus the calculation of the Chern-Simons invariant
reduces to the representation theory of the fundamental group into
$SU(2)$. In \cite{FinSte:90} the authors define a further invariant\[
\tau(\Sigma)=\min\left\{ CS(\alpha)|\:\alpha:\pi_{1}(\Sigma)\rightarrow SU(2)\right\} \]
for the 3-manifold $\Sigma$. This invariants fulfills the relation\[
\tau(\Sigma)=\frac{1}{8\pi^{2}}\int\limits _{\Sigma\times\mathbb{R}}tr(F_{A}\wedge F_{A})\]
which is the minimum of the Yang-Mills action \[
\left|\frac{1}{8\pi^{2}}\int\limits _{\Sigma\times\mathbb{R}}tr(F_{A}\wedge F_{A})\right|\leq\frac{1}{8\pi^{2}}\int\limits _{\Sigma\times\mathbb{R}}tr(F_{A}\wedge*F_{A})\]
i.e. the solutions of the equation $F_{A}=\pm*F_{A}$. Thus the invariant
$\tau(\Sigma)$ of $\Sigma$ corresponds to the self-dual and anti-self-dual
solutions on $\Sigma\times\mathbb{R}$, respectively. Or the invariant
$\tau(\Sigma)$ is the Chern-Simons invariant for the Levi-Civita
connection.


\end{document}